\documentclass{emulateapj}
\usepackage{lscape} 
\usepackage{natbib}
\newcommand \lsol{L$_{\odot}$}
\newcommand \msol{M$_{\odot}$}

\newfont{\rten}{cmr10} 
\def \arcdeg{\hbox{$^\circ$}}

\begin{document}


\title{Locating planetesimals belts in the multiple-planet systems HD 128311, HD 202206, HD 82943 and HR 8799}

\author{Amaya Moro-Mart\'in\altaffilmark{1,2}, 
Renu Malhotra\altaffilmark{3}, 
Geoffrey Bryden\altaffilmark{4}, 
George H. Rieke\altaffilmark{5}
Kate Y. L. Su\altaffilmark{5},
Charles A. Beichman\altaffilmark{6}, 
Samantha M. Lawler\altaffilmark{7}}

 
\altaffiltext{1}{Department of Astrophysics, Center for Astrobiology (CSIC-INTA),  Ctra. de Ajalvir, km 4, Torrej\'on de Ardoz, 28850, Madrid,  Spain}
\altaffiltext{2}{Department of Astrophysical Sciences, Princeton University, Peyton Hall, Ivy Lane, Princeton, NJ 08544, USA}
\altaffiltext{3}{Department of Planetary Sciences, University of Arizona, 1629 E. University Boulevard, Tucson, AZ 85721, USA}
\altaffiltext{4}{Jet Propulsion Laboratory, 4800 Oak Grove Drive, Pasadena, CA 91109, USA}
\altaffiltext{5}{Steward Observatory, University of Arizona,  933 North Cherry Ave, Tucson, AZ 85721, USA}
\altaffiltext{6}{NASA Exoplanet Science Institute, California Institute of Technology,  Pasadena, CA 91125, USA} 
\altaffiltext{7}{Department of Physics and Astronomy, University of British Columbia, 6224 Agricultural Road, Vancouver, BC V6T 1Z1, Canada}

\begin{abstract}
In addition to the Sun, six other stars are known to harbor multiple planets and debris disks: HD 69830, HD 38529, HD 128311, HD 202206, HD 82943 and HR 8799. In this paper we set constraints on the location of the dust-producing planetesimals around the latter four systems. We use a radiative transfer model to analyze the spectral energy distributions of the dust disks (including two new $Spitzer$ $IRS$ spectra presented in this paper), and a dynamical model to assess the long-term stability of the planetesimals'  orbits.  As members of a small group of stars that show evidence of harboring a  multiple planets and planetesimals, their study can help us learn  about the diversity of planetary systems. 
\end{abstract}


\section{Introduction}

Surveys with the {\it Spitzer} Space Telescope have been spectacularly successful at identifying infrared excess emission associated with planetary debris disks around A- through K-type stars. The excesses at 70 $\mu$m are associated with cool dust located at distances from the stars analogous to the position of the Kuiper Belt (KB) in the solar system (Moro-Mart\'in et al. 2008), although much larger numbers of objects must lie in these exo-Kuiper belts to account for the detected level of emission (Trilling et al. 2008; Hillenbrand et al. 2008; Carpenter et al. 2009). The most sensitive studies find 70 $\mu$m excesses around $\sim$ 15-20\% of mature solar-type stars over the entire 10 Myr to 10 Gyr age range (Trilling et al. 2008). The excesses at 24$\mu$m are generally associated with warmer dust and disappear relatively quickly as the host star ages; about 30\% of the stars of the Pleiades age ($\sim$ 120 Myr) show excess emission (Sierchio et al. 2010) whereas by the age of Praesepe ($\sim$ 600 Myr), the 24 $\mu$m excesses have almost completely disappeared (Meyer et al. 2008; G\'asp\'ar et al. 2009). Because the expected lifetimes of the debris dust grains are much shorter than the ages of the stars, it is inferred that the dust originates from collisional activity in reservoirs of planetesimals left over from the planet formation process (hence the term {\it debris} dust). To sustain the dust production, it is necessary that large planetesimals (1000 km-sized) or unseen planets stir the planetesimals so they continue to collide with each other. Planets are also responsible for constraining the planetesimals in some zones and clearing them from others, thus determining much of the structure of the debris system. The way the patterns of debris disks activity decay with age is consistent with the expectation that the inner zones of a planetary system have relatively short dynamical time scales, whereas dynamical activity unfolds slowly at the distance of the Kuiper Belt. 

A highlight of the recent surveys is the first detection of debris disks around stars with planets (Beichman et al. 2005; Moro-Mart\'in et al. 2007a). These planets orbit within several AU of their parent star, whereas the cold dust emitting the observed far-IR radiation generally resides tens of AU away. Despite the separation between the dust and planets, it is still possible  for the planet to shape the structure of  the dust (and planetesimals) disk. In the multiple-planet system HD 38529, for example, secular resonances excited by planets at 0.13 and 3.7 AU create regions at tens of AUs that are unstable for orbiting planetesimals (Moro-Mart\'in et al. 2007b). 

Several additional systems have been identified as having both multiple planets (capable of exciting secular resonances) and orbiting debris (indicating the presence of planetesimals; Bryden et al. 2009, Su et al. 2009). Presumably there are many more stars with debris-disk excesses that also harbor multiple but undiscovered planets. Detailed studies of the known examples can reveal aspects of their behavior that help us understand the diversity of
planetary systems. In this paper we study four of these systems: HD 128311, HD 202206, HD 82943 and HR 8799. In Section \ref{obs} we describe the planet and debris dust detections for each one of these systems, and present new $Spitzer$ $IRS$  detections  for the debris disks around HD 202206 and HD 82943. In Section \ref{sed} we use a radiative transfer model to identify the range of parameters (dust mass and dust location) that would fit the observed spectral energy distribution (SED). Due to the high fractional luminosity, the grain-grain collisional time-scale is shorter than the Poynting-Robertson (P-R) time-scale for all these systems and therefore we expect the dust to trace the location of the dust-producing planetesimals.  In Section \ref{dyn}, we use a dynamical model to assess the long-term orbital stability of the putative dust-producing planetesimals, taking into account the effect of secular resonances.  Putting together the results from the SED and dynamical analysis, in Section \ref{loc} we discuss the potential location of the dust-producing planetesimal belts. 
\section{Debris disks detected in multiple-planet systems}
\label{obs}

\begin{figure*}
\begin{center}
\includegraphics[scale=0.6,angle=0]{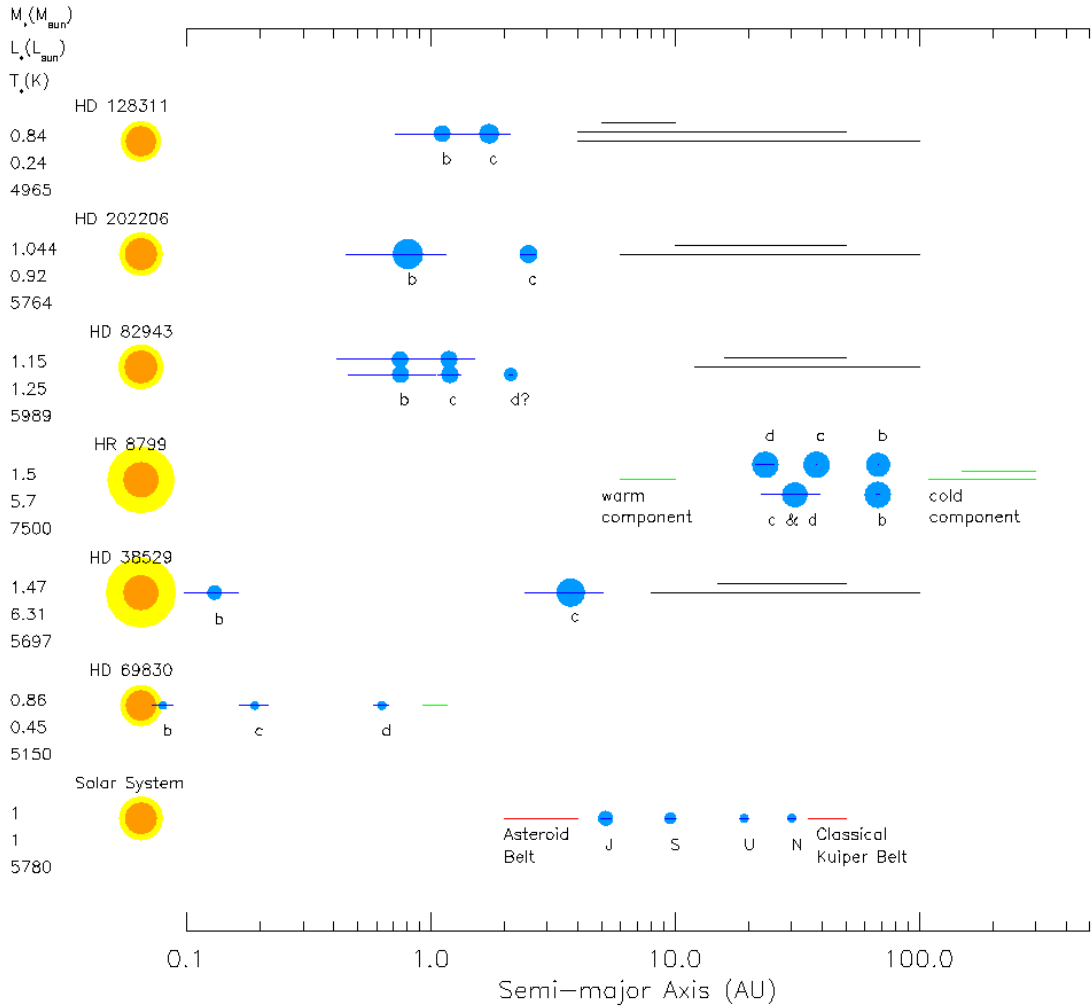}
\end{center}
\caption{Schematic representation of the seven planetary systems known to harbor multiple planets and dust-producing planetesimals. The stars are represented by the orange/yellow circles, with the stellar mass, luminosity and effective temperature labeled to the left.  The size of the orange circle is proportional to the cube root of the stellar mass, while the size of the yellow circle is proportional to the stellar luminosity. The planets are represented by blue symbols with sizes proportional to the cube root of the planet mass. The thin blue lines extend from periastron to apoastron. For a given system, there is a range of planetary configurations that can fit the observations  (see the planets' orbital elements  in Table \ref{planetparam}); many of these possible configurations would look very similar to each other under this schematic representation, with the exception of HD 82943 and HR 8799; for each of these stars we show two possible planetary configurations that significantly differ from each other. The inferred location of the dust-producing planetesimals are represented by the black and the green lines. Each line corresponds to a possible solution of a single-component disk, showing the degeneracy of the problem (exceptions are HR 8799, where warm and cold dusty disk components are inferred to exist from the SED, and the Sun, with the Asteroid Belt and the Kuiper Belt). The black lines correspond to solutions that assume that the dust is composed of 10 $\mu$m-sized grains, while the green lines correspond to models that assume a grain size distribution (see Table \ref{multidisks} for details); the asteroid and Kuiper belts are shown in red, the only two planetesimal belts that have been directly detected. 
} 
\label{arch}
\end{figure*}

Including the solar system, 19 planetary systems are currently known to have both orbiting debris and planets (Bryden et al. 2009), and while the majority are single-planet systems, seven of them are known to harbor multiple planets. These systems are described in Figure \ref{arch} and Table  \ref{multidisks}. Figure \ref{arch} also shows the locations of the planetesimal belts  derived either from work in the literature or in this paper. Our estimates arise from SED fitting together with dynamical  models that study the effects imposed by the planets on the stability of the planetesimals.  In this section we describe the infrared excesses, the planet detections, and the dynamical models for the planet orbits.

For these multi-planet systems, the planet-dust interaction has previously been studied in the case of the solar system,  HD 38529 (Moro-Mart\'{\i}n et al. 2007b) and HD 69830 (Lovis et al. 2006, Lisse et al. 2007). In this paper we study the remaining four systems: HD 128311, HD 202206, HD 82943 and HR 8799. 

\begin{deluxetable*}{l|ccccccc}
\tablewidth{0pc}
\tablecaption{{\it Spitzer} Photometry \label{spitphotom}}
\tablehead{
\colhead{Star} & \multicolumn{7}{c}{Wavelength ($\mu$m)} \\
\colhead{} &
\colhead{   13.20 $^a$ }	&  
\colhead{   19 $^a$ }	&  
\colhead{   23.68 }	&   
\colhead{   25 $^a$ }	&  
\colhead{   32.5 $^a$} 	&   
\colhead{   71.42$^b$}  	&   
\colhead{   155.9 }   	
}
\startdata
 HD 128311 &
 192.45  $\pm$   15.62	&
&
 60.0    $\pm$   1.2   	&
&
 32.95   $\pm$   4.82  	&
 23.5    $\pm$   3.3 $^c$   	&
 50.0    $\pm$   35.0	\\
 HD 202206 &
   57.57$\pm$3.80   	&
    28.94$\pm$1.95   	&
   N/A &
   17.22$\pm$1.16   	&
   14.23$\pm$0.83  $^c$  	&
 32.0  $\pm$  3.1 $^c$  	&
 N/A   	\\
  HD 82943 &
   201.10$\pm$10.87   	&
   98.29$\pm$6.99	      &
   66.0$\pm$1.3 &
	&
   46.98$\pm$3.27   	&	
133$\pm$5$^c$    &
 N/A   	\\
\enddata
\tablecomments{Fluxes are given in mJy. The uncertainties are 1-$\sigma$.}
\tablenotetext{a}{Synthetic fluxes calculated from the $IRS$ spectrum using a square filter over the following wavelength ranges: 13.2 $\mu$m  (12.4--14 $\mu$m), 19 $\mu$m (18--20 $\mu$m),   25 $\mu$m  (24--26 $\mu$m), and 32.5 $\mu$m  (30--35 $\mu$m). The uncertainties in this case are calculated from the standard deviation of the flux within the wavelength bin.}
\tablenotetext{b}{The 70 $\mu$m flux is color corrected assuming a dust temperature of 50 K and a correction factor of 0.893 (from the {\it Spitzer} Users Manual).}
\tablenotetext{c}{Measured flux is significantly above that expected from the stellar photosphere.}
\end{deluxetable*}

\subsection{Observations of HD 128311}
\label{obs_HD128311}
HD 128311 is a K0 star located at 16.57 pc, with T$_{eff}$ = 4965 K,  M$_*$ = 0.84 \msol~and a metallicity of [Fe/H] = -0.04 (Saffe et al. 2008). King et al. (2003) identified this star as a possible member of the UMa moving group, suggesting an age of $\sim$ 500 Myr, consistent with the estimate of 390 to 410 Myr by Saffe et al. (2005). The chromospheric activity index reported by Gray et al. (2003) indicates an age of $\sim$ 560 Myr, using the calibration of Mamajek \& Hillenbrand (2008). Barnes (2007) found an age of 350 Myr from gyrochronology. All of these determinations are consistent within their errors with an age of 500 Myr, which also lies at the lower end of the range from isochrone fitting (Valenti \& Fischer 2005).

 \subsubsection{Debris dust detections}
\begin{figure}
\begin{center}
\includegraphics[scale=0.3,angle=90]{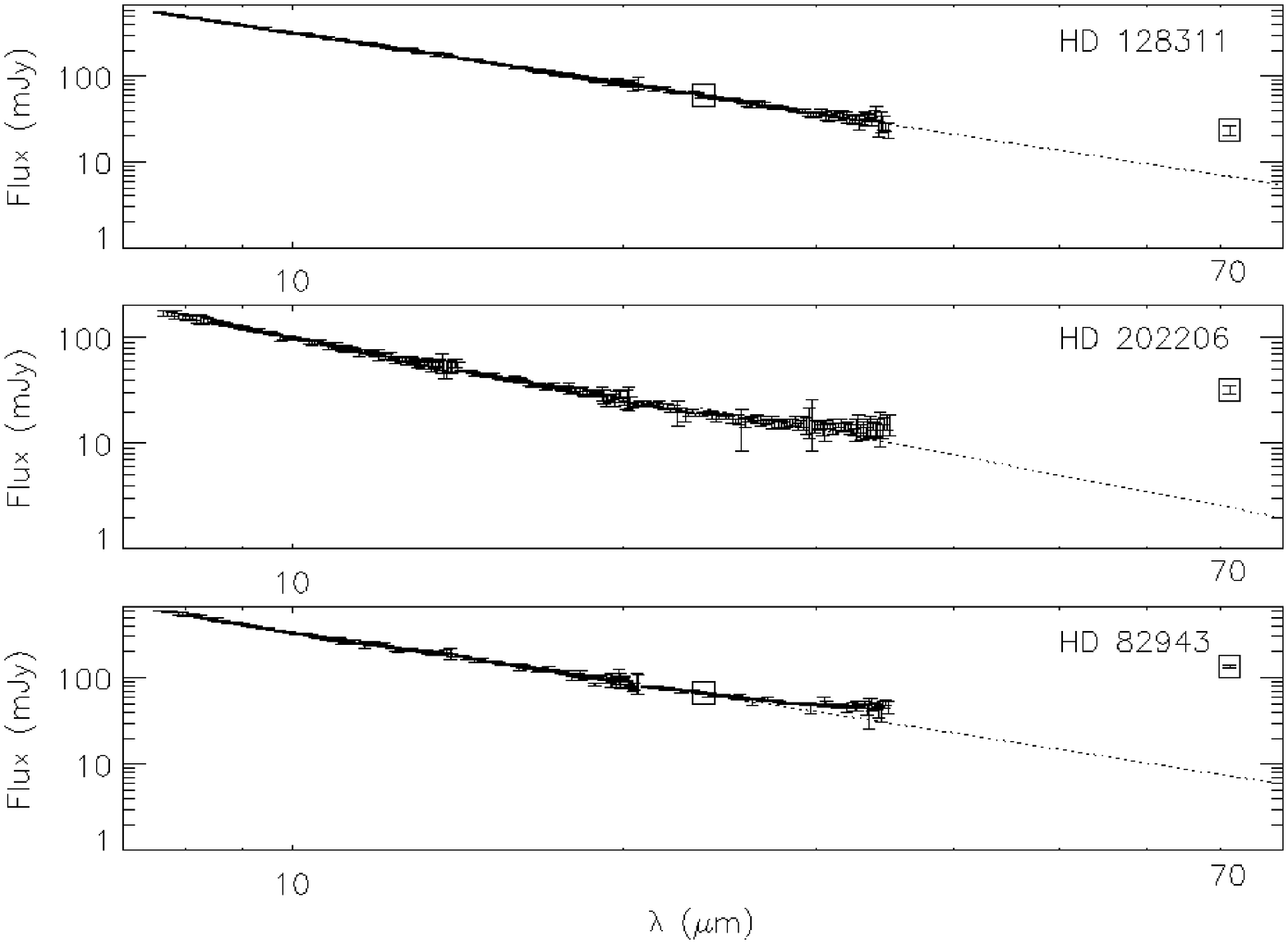}
\end{center}
\caption{Spitzer observations of HD 128311, HD 202206 and HD 82943. The {\it IRS} spectrum is shown as a continuous line, while squares are the {\it MIPS} photometric points; in both cases the error bars correspond to 1-$\sigma$ uncertainties. The dotted line is the stellar photosphere (approximated as a blackbody). 
} 
\label{obs_3stars}
\end{figure}

HD 128311 was observed by a {\it Spitzer} Guaranteed Time Observation program specifically targeting planet-bearing stars. The {\it Spitzer} observations are shown in Figure \ref{obs_3stars} and Table \ref{spitphotom}. The star has a strong excess at 70 $\mu$m (Beichman et al. 2005; Trilling et al. 2008), and no excess was detected with {\it MIPS} at 24 $\mu$m (Trilling et al. 2008). No excess was seen in the {\it IRS} spectrum at 5--35 $\mu$m (Beichman et al. 2006). [HD 128311 has a {\it MIPS} 24 $\mu$m flux of 60$\pm$1.2 mJy (1-$\sigma$ uncertainty), somewhat lower than but consistent with the flux expected from a K-band extrapolation (64 mJy)]. 

Excesses at 70 $\mu$m are relatively common. Hillenbrand et al. (2008) and Carpenter et al. (2009) report them for 6--10\% of the 328 FGK stars in the FEPS sample, while the generally deeper (relative to the photospheric level) measurements of Trilling et al. (2008) find them in 16.4 $\pm$ 2.9 \% of the FGK stars in their sample. However, HD 128311 may be exceptional in not having an excess at 33 $\mu$m, because only 3 out of 152 FGKM stars surveyed by Lawler et al. (2009) showed a 70 $\mu$m excess with no corresponding 33 $\mu$m. 
 
\subsubsection{Planet detections}
Radial velocity monitoring of HD 128311 has led to the discovery of two planets.  We found that the orbital solution in Vogt et al. (2005) is not stable. In this paper we have considered the following two orbital solutions (see planetary parameters in Table \ref{planetparam}):
\begin{itemize}
\item {\bf Fit A1} is the solution in Butler et al. (2006), with the 2 planets near the 2:1 mean motion resonance (MMR).
\item {\bf Fit A2} (labeled as fit I in Gozdzlewski \& Konacki 2006) corresponds to the best fit, co-planar solution lying at the border of an island of stability related to the corotation of apsides, with the planets also in the 2:1 MMR. 
\end{itemize}
Gozdzlewski and Konacki (2006) suggested that the radial velocity data can also be explained by two planets in a non-co-planar 1:1 MMR. In the present paper, we do not consider this somewhat exotic solution, as we assume that the protoplanetary flattened disk results in a system where the planets and the dust-producing planetesimals are on the same plane. 

Direct planet searches in the outer planetary system with the MMT (Biller et al. 2007) and VLT/NACO (Eggenberger et al. 2007) have led to non-detections, however, they can set no constraints  on the presence of planets smaller than 10~M$_{Jup}$. 

\subsection{Observations of HD 202206}
\label{obs_HD202206}
HD 202206 is a metal-rich G6 V star located at 46.3 pc, with T$_{eff}$ = 5765 K,  M$_*$ = 1.15 \msol, a metallicity of [Fe/H] = 0.37$\pm$0.07  and a stellar age of 5.6$\pm$1.2 Gyr (Udry et al. 2001); Saffe et al. 2005 estimates an age of 4.2 Gyr. 

\subsubsection{Debris dust detections}
HD 202206 was observed by a {\it Spitzer} Cycle 4 General Observer program targeting planet-bearing stars that were missed by the earlier Guaranteed Time Observations.  The {\it Spitzer} observations are shown in Figure \ref{obs_3stars}; Table \ref{spitphotom} summarizes the (synthetic-)photometry.
A strong excess was detected at 70 $\mu$m, a factor of 13 above the stellar photosphere (Bryden et al. 2009). In this paper we present new {\it IRS} observations which show emission in excess of the stellar photosphere for $\lambda > 25$ $\mu$m (see also Dodson-Robinson 2010, in preparation). No 24 $\mu$m observations were made. We use the new {\it IRS} spectrum to calculate fluxes in four synthetic bands. 

\subsubsection{Planet detections}

From radial velocity monitoring, Correia et al. (2005) concluded there are two planets in this system: a very massive 17.5 M$_{Jup}$ inner body and a less massive 2.4 M$_{Jup}$ outer planet. This system is particularly interesting because of the high mass of the innermost body (a brown dwarf). If it formed in the circumstellar protoplanetary disk, its existence would imply that such disks can be extremely massive, whereas if it formed like a stellar companion this system would be the only known example of a circumbinary planet orbit. In the latter situation, Nelson (2003) predicted that the interaction of the outermost planet with the viscous circumbinary disk could have been responsible for its inward migration and resonance trapping. 

From stability considerations, Correia et al. (2005) suggested that the system is in an island of stability around the 5:1 mean motion resonance (MMR), with $\it{M}$sin$\it{i}$ of  17.428 M$_{Jup}$ and 2.436 M$_{Jup}$, semi-major axes of  0.830 AU and 2.542 AU, and eccentricities of 0.435 and 0.267, for HD 202206 b and c, respectively. This solution has a ($\chi_{\nu}$$^{2})^{1/2}$ of 1.67 and is stable over a 5 Gyr timescale.  However,  using n-body analysis  that takes into account stability considerations and is well suited for multi-planet systems in low-order MMRs, Gozdziewski et al. (2006) concluded that the solution by Correia et al. (2005) is a local minimum. Assuming a co-planar system, their best fit (with r.m.s = 9.98 m/s and ($\chi_{\nu}$$^{2})^{1/2}$ = 1.53) has $\it{M}$sin$\it{i}$ of 17.624  M$_{Jup}$ and 2.421 M$_{Jup}$, semi-major axes of  0.831 AU and 2.701 AU, and eccentricities of  0.433 and 0.255, for HD 202206 b and c, respectively. [Gozdziewski et al. (2006) noted that an unexpected non-co-planar solution is also possible, where the eccentricity of the innermost largest planet varies with larger amplitude than that of the outermost planet, while the inclination of the latter can assume almost any value]. In a more recent paper, Couetdic et al. (2009) used frequency map analysis and updated radial velocity data to explore the long term stability of a wide  range of possible orbital solutions. Adopting an updated stellar mass of 1.044 M$_{\odot}$, they found acceptable co-planar configurations  for inclinations between 30$\degr$--90$\degr$ with respect to the line  of sight and favored an edge-on co-planar solution in which the bodies are in the 5:1 MMR (with ($\chi_{\nu}$$^{2})^{1/2}$ = 1.4136). This solution has $\it{M}$sin$\it{i}$ of 16.59  M$_{Jup}$ and 2.179 M$_{Jup}$, semi-major axes of  0.8053 AU and 2.49 AU, and eccentricities of   0.431 and 0.104, for HD 202206 b and c, respectively. It is stable for more than 5 Gyr and differs from the solutions above in that the eccentricity of planet c is lower, allowing more stable regions outside the resonances. Couetdic et al. (2009) explored the stability of test particles in the above configuration identifying two possible niches for stability: $a <$ 0.12 and $a >$ 6.5 AU (for an integration time of 16000 years).  Couetdic et al. (2009) noted that the solution above has a high resonant mode amplitude that will likely be dampened by dissipative forces. 

\begin{itemize}
\item {\bf Fit B1} in Table \ref{planetparam} is the edge-on co-planar configuration favored by Couetdic et al. (2009); this solution is close to the center of libration and has ($\chi_{\nu}$$^{2})^{1/2}$ = 1.553. 
\end{itemize}

The radial velocity data acquired so far cannot exclude the presence of a Neptune-sized planet between 0.06 AU--0.12 AU, or a 10 M$_{\oplus}$ planet between 0.02 AU--0.12 AU, or a 0.5 M$_{Jup}$ mass planet outside 6.5 AU. No planetary companions have been found by direct imaging (a source detected by Chauvin et al. 2006 was a background object). 

\subsection{Observations of HD 82943}
\label{obs_HD82943}
HD 82943 is a  G0 V star located at 27.46 pc, with T$_{eff}$ = 5989 K,  M$_*$ = 1.15 \msol~and [Fe/H] = 0.26  (Sousa et al. 2008). Consistent measurements of the chromospheric activity have been reported by Wright et al. (2004), Gray et al. (2006), and Saffe et al. (2005). Using the calibration of Mamajek and Hillenbrand (2008), they imply an age of 5 Gyr.

\subsubsection{Debris dust detections}

HD 82943 was observed by a {\it Spitzer} Guaranteed Time Observation program targeting nearby late F, G, and early K stars. The {\it Spitzer} observations are shown in Figure \ref{obs_3stars}; Table \ref{spitphotom} summarizes the (synthetic-)photometry. Trilling et al. (2008) found a strong excess emission at 70 $\mu$m and no excess at 24 $\mu$m, deriving a minimum characteristic dust temperature of 69 K (corresponding to 22 AU if assuming blackbody grains), and a fractional luminosity of L$_{dust}$/L$_{star}$ $\sim$ 10$^{-4}$. In this paper we present new {\it IRS} observations that show how the spectrum rises above the stellar photosphere for $\lambda > 26$ $\mu$m. 

\subsubsection{Planet detections}
From radial velocity observations, Mayor et al. (2004) announced the presence of two planets around HD 82943. Ferraz-Mello et al. (2005) found that the orbital solution in the discovery paper is unstable and that the radial velocity observations can be fitted with a stable, co-planar solution in which the planets are locked in the 2:1 MMR. Gozdziewski \& Konacki (2006) revisited the orbital solution modeling data from CORALIE (Mayor et al. 2004) and from Keck-HIRES (Lee et al. 2006), confirming that the co-planar 2:1 MMR solution fits the observations. They also found a stable, non-co-planar solution in a 1:1 MMR, but as we mentioned above in this paper we only consider co-planar configurations. The dynamical maps in  Gozdziewski \& Konacki (2006) revealed that near their unstable, best fit solution there are two narrow islands of stability associated with the 2:1 MMR (see orbital elements in Table \ref{planetparam}): 
\begin{itemize}
\item  {\bf Fit C1} (labeled fit V in Gozdziewski \& Konacki 2006) lies in one of the two islands and corresponds to their best fit, rigorously stable, two-planet solution, characterized by co-rotation of the apsidal lines. [We found that the second solution they identify in their Figure 9 becomes unstable after 27.8 Myr and therefore we do not consider it in this paper; this solution lies at the very edge of the second island of stability of the 2:1 MMR, where slight differences in the initial conditions can result in a very different dynamical evolution.]
\end{itemize}
Gozdziewski \& Konacki (2006) and Beauge et al. (2008) discussed the inadequacy of the 2-body solution and speculated on the presence of a third planet to account for the large r.m.s. and the lack of convergence of some of the orbital elements (eccentricity and longitude of periapsis) as the number of radial velocity data points increases.  In this paper, we have also considered the following stable, 3-planet, co-planar solutions (see orbital elements in Table \ref{planetparam}):  
\begin{itemize}
\item {\bf Fit C2} is the solution in Gozdziewski \& Konacki (2006), in which planets b and c are in the 2:1 MMR while the outermost planet d is in a low eccentricity non-resonant orbit. 
\item {\bf Fit C3} is the solution in  Beauge et al. (2008), in which the planets are in a Laplace 1d:2c:4b  MMR in a double asymmetric apsidal corotation resonance. 
\end{itemize}

\subsection{Observations of HR 8799}
\label{obs_HR8799}
HR 8799 is a metal-poor A5 V star located at 39.4 pc, with T$_{eff}$ = 7500 K,  M$_*$ = 1.5 \msol~and a metallicity of [Fe/H] = -0.55. Marois et al. (2008) estimated a stellar age of 30--160 Myr; Moya et al. (2010) argued that the age is still unconstrained, with astroseismology analysis favoring an older age of $\sim$ 1 Gyr, but requiring a better determination of the rotation velocity of the star.

\subsubsection{Debris dust detections}

HR 8799 was observed by {\it Spitzer} under programs 50175 and 530; images and an SED analysis were presented in Su et al. (2009). Deep {\it IRS} and {\it MIPS} observations (including the MIPS SED mode from 55--95 $\mu$m  -- a wavelength range critical to constrain the shape of the SED), revealed a spatially resolved disk at 24 and 70 $\mu$m, whose outer boundary can be traced beyond 1000 AU. The excess dust emission can be detected in the IRS spectra  starting bellow 20 $\mu$m (less deep {\it IRS} observations are discussed in Chen et al. 2009) and also at 160 $\mu$m. Unresolved debris disk detections  exist at  60 and 90 $\mu$m (with {\it ISO}; Mo\'or et al. 2006), at 850 $\mu$m (with  JCMT/SCUBA; Williams \& Andrews 2006) and at 1.2 mm (Sylvester et al. 1996). Preliminary analysis shows that the debris disk is spatially resolved with APEX/LABOCA at 870 $\mu$m (Kalas et al. 2010). 
 
\subsubsection{Planet detections}
Three distant planets have been detected by direct imaging at projected separations of 68 AU, 38 AU and 24 AU (for HR 8799 b, c and d, respectively -- Marois et al. 2008; Lafreniere et al. 2009); the planets' long orbital periods  and the short baseline of the observations make the determination of the orbital parameters still uncertain. It is possible to estimate the masses of the directly detected planets from their luminosities by using evolutionary cooling models (assuming the age of the system is known).  However, current models are discrepant and, in the case of HR 8799, the age of the system is uncertain, making the planets' masses poorly constrained;  lower limits to the masses of HR 8799 b, c and d are in the ranges of 5--11, 7--13 and 7--13 M$_{Jup}$, with nominal masses of 7, 10 and 10 M$_{Jup}$, respectively (Marois et al. 2008; Reidemeister et al. 2009). 

Fabrycky \& Murray-Clay (2010) found that many orbital solutions are unstable, including the face-on circular configuration (for any range of reasonable planet masses).  
They tried to determine the system parameters by fitting simultaneously the astrometry (positions and proper motions), the planets' luminosities (that constrain the range of possible planets' masses), and the requirement that the system is stable for at least 30-160 Myr.  

\begin{itemize}
\item {\bf Fit D1} is the face-on, co-planar solution identified by Fabrycky \& Murray-Clay (2010) that is  stable for $>$ 160 Myr even if the planets' masses are up to 1.9 times their nominal values; in this solution the planets are locked into a Laplace 1d:2c:4b MMR, allowing them to avoid close encounters (see orbital elements in Table \ref{planetparam}). 
\end{itemize}

Reidemeister et al. (2009) have also explored the stability of the HR 8799 system in the case of co-planar and initially circular orbits, where the orbital plane is allowed to have a range of inclinations ($I$ = 0--40$\degr$) and orientations ($\Omega$' = 0--180$\degr$). Here, $I$ is the angle between the angular momentum vector and the vector toward the observer, and $\Omega$' is the angle between the North direction and the line of nodes (measured toward the East -- see Figure 1 in Reidemeister et al. 2009). They integrated a grid of models over 100 Myr and identified as stable the configurations where the planets  did not suffer close encounters during that period; these configurations (all co-planar) have $I > $ 20$\degr$ and $\Omega$' = 0--50$\degr$.  We take into account the following considerations:
\begin{itemize} 
\item Figure 8 in Reidemeister et al. (2009), showing the models that fulfill the stability criteria (i.e. no close encounters during 100 Myr).
\item Figure 9 in Reidemeister et al. (2009), showing the models that fit the observed astrometry.  
\item Inclination constrains of the equatorial plane of the star derived from the comparison of the measured rotational velocity $v$$\cdot$sin$i$ and the expected rotational velocity for an A5 star:  $I$ = 7--22$\degr$ (Lafreniere et al. 2009). 
\item Inclination constraints from astrometry: (1) assuming circular, co-planar orbits and fixing the stellar mass to 1.5 M$_{\odot}$, but letting the orientation of the orbit float: $I \sim 20\degr$ (Fabrycky \& Murray-Clay 2010); (2) from observations with a 10 year baseline (using NICMOS archival data where the planets are detected): $I$ = 13--23$\degr$ (Lafreniere et al. 2009).
\item Inclination constraints from the debris disk morphology: (1) from Spitzer spatially resolved observations: $I <$ 25$\degr$ (Su et al. 2009); (2) from preliminary analysis of 870 $\mu$m observations: $I >$ 20$\degr$ (Paul Kalas, private comm.). 
\end{itemize}
Based on the above considerations, we have adopted the following two additional non-face-on, co-planar configurations (see orbital elements in Table \ref{planetparam}): 
\begin{itemize} 
\item {\bf Fit D2} with ($I = $ 25$\degr$, $\Omega$' = 20$\degr$). 
\item {\bf Fit D3} with ($I = $ 20$\degr$, $\Omega$' = 45$\degr$). 
\end{itemize}

As for the planets' masses, we initially considered: (1) the nominal masses of  7, 10 and 10 M$_{Jup}$ for HR 8799 b, c and d, respectively; and (2) the low mass case with 5, 7 and 7 M$_{Jup}$ for HR 8799 b, c and d, respectively. However, we found that when adopting nominal planet masses, the configurations D2 and D3 were unstable. 

Based on the location of HR 8799 b in a pre-discovery 2007 image  (Metchev et al. 2009),  Fabrycky \& Murray-Clay (2010) ruled out stable co-planar configurations in which the innermost planet is very eccentric ($e >$ 0.95) or retrograde. They also found very non-co-planar (non-resonant) configurations that are stable but we will not consider those because: (1) they imply we are observing the system at a special time; and (2) it is more reasonable to assume that the planets and the planetesimals that produce the dust are co-planar because they formed out of a flattened disk. 

In a recent study of HR 8799, Gozdziewski \& Migaszewski (2009) used a self-consistent, n-body analysis that takes into account stability considerations to carry out a quasi-global search for stable, co-planar planetary configurations that can fit the astrometric data and the astrophysical mass constraints for the star and the planets. In their study, the planets and stellar masses are free parameters and are allowed to vary within their 1-$\sigma$ range, with M$_b$ = 7$^{+4}_{-2}$ M$_{Jup}$, M$_c$ = 10$\pm$3 M$_{Jup}$, M$_d$ = 10$\pm$3 M$_{Jup}$ and M$_*$ = 1.5$\pm$0.3 M$_{\odot}$. As in Fabrycky \& Murray-Clay (2010) and Reidemeister et al. (2009), the goal of their study  is to use the requirement of stability to identify long-lived configurations that may lie close to unstable best-fitting solutions, mitigating the errors introduce by the short baseline of the observations.  Gozdziewski \& Migaszewski (2009) found two long-term, stable,  non-face-on, co-planar solutions (see orbital elements in Table \ref{planetparam}):
\begin{itemize}
\item {\bf Fit D4} ($I = $ 15.5$\degr$, $\Omega$' = 11.2$\degr$): inside a small island of regular motion; the planets are in the 1d:2c:4b MMR; the solution becomes unstable after 400 Myr. 
\item {\bf Fit D5} ($I = $ 11.4$\degr$, $\Omega$' = 357.2$\degr$): also inside a small island of stable motion; the inner two planets are in the 1c:1d MMR, and the outermost planet is in a narrow island close to the (1c:1d):3b MMR; the inner planets are stable for $>$ 3 Gyr, while the 3-planet system has a regular motion only during the first 600 Myr. 
\end{itemize}

Table \ref{planetparam} lists the orbital elements of the five  planetary configurations (Fits D1--D5) we have adopted for our test particle simulations of HR 8799 in Section \ref{dyn_HR8799}. Our numerical runs last 160 Myr (the upper limit to stellar age estimated by Marois et al. 2008); there is no guarantee that the systems will remain stable on longer time-scales. 

\begin{deluxetable*}{lcccccccccc}
\tablewidth{0pc}
\tablecaption{Planetary and stellar parameters for the dynamical simulations\label{planetparam}}
\tablehead{
\colhead{Planet} &
\colhead{$M_{*}$} &
\colhead{($I, \Omega$')} &
\colhead{$M_{planet}$} &
\colhead{$a$} &
\colhead{$e$} &
\colhead{$i$} &
\colhead{$\Omega$} &
\colhead{$\omega$} &
\colhead{$M$} &
\colhead{Ref} \\
\colhead{} &
\colhead{(M$_{\odot}$)} &
\colhead{(deg,deg)} &
\colhead{(M$_{Jup}$)} &
\colhead{(AU)} &
\colhead{} &
\colhead{(rad)} &
\colhead{(rad)} &
\colhead{(rad)} &
\colhead{(rad)} &
\colhead{}} 
\startdata
HD 128311					& 0.84 	& &    	&	&	&	&	&   &	&   B06 \\
Fit A1						& &  (90,\nodata) &   &	&	&	&	&   &	&    \\
~~~~~~~~~~~~~~~~b  		& & &		2.19		& 	1.10 	&	0.25 	& 0		& 0		& 	1.93732  	& 2.89	 &	 \\
~~~~~~~~~~~~~~~~c		& & &    	3.22  	 	&	1.76 	&	0.17 	& 0		& 0		& 	3.49066  	& 0.068 &	 \\
\hline\\
HD 128311					& 0.84 & & &	& & &	&   &	&   G06 \\
Fit A2						& &  (90,\nodata)	&   &	&	&	&	&   &	&    \\
~~~~~~~~~~~~~~~~b		& & &		1.606		& 	1.112 	&	0.359 	& 0		& 0		& 	1.24931   		& 4.74241	 &	 \\
~~~~~~~~~~~~~~~~c		& & &    	3.178  	 	&	1.732 	&	0.214 	& 0		& 0		& 	0.221831  	& 3.32014 &	 \\
\hline
\hline\\
HD 202206	 		& 	1.044 	& & &	&	&	&	&   &	& C09 \\
Fit B1	 			& & (90,\nodata) & &	&	&	&	&   &	& \\
~~~~~~~~~~~~~~~~b		& & & 	 16.59		&	0.8050		&	0.439		& 0 	&	0		& 2.82412		& 1.35158 \\
~~~~~~~~~~~~~~~~c		& & &  	 2.179		&	2.5113  	&	0.071		& 0 	&	0		& 1.36467		& 2.95153	 \\
\hline
\hline\\
HD 82943	 			& 	1.15 	& & &	&	&	&	&   &	& G06 \\
Fit C1	 			& & (90,\nodata) & &	&	&	&	&   &	& \\
~~~~~~~~~~~~~~~~b			& & & 	 1.461		&	0.748	&	0.448		& 0 	&	0		& 2.21343		& 6.26975 \\
~~~~~~~~~~~~~~~~c			& & &  	 1.728		&	1.186  	&	0.268		& 0 	&	0		& 2.41466		& 5.87914	 \\
\hline\\
HD 82943	 			& 	1.15 	& & &	&	&	&	&   &	& G06 \\
Fit C2	 			& & (90,\nodata) & &	&	&	&	&   &	& \\
~~~~~~~~~~~~~~~~b			& & & 	 1.679		&	0.751	&	0.386		& 0 	&	0		& 2.06088		& 0.04625 \\
~~~~~~~~~~~~~~~~c			& & &  	 1.867		&	1.197  	&	0.110		& 0 	&	0		& 2.52148		& 6.02557	 \\
~~~~~~~~~~~~~~~~d			& & &  	 0.487		&	2.125  	&	0.018		& 0 	&	0		& 2.00032		& 1.39207	 \\
\hline\\
HD 82943	 			& 	1.15 	& & &	&	&	&	&   &	& B08 \\
Fit C3	 			& & (90,\nodata) & &	&	&	&	&   &	& \\
~~~~~~~~~~~~~~~~b			& & & 	 1.703		&	0.745	&	0.361		& 0 	&	0		& 2.30942		& 6.11127 \\
~~~~~~~~~~~~~~~~c			& & &  	 1.747		&	1.200  	&	0.190		& 0 	&	0		& 3.35121		& 5.20335	 \\
~~~~~~~~~~~~~~~~d			& & &  	 0.351		&	1.912  	&	0.078		& 0 	&	0		& 2.0279		& 0.482234	 \\
\hline
\hline\\
HR 8799			&	1.5 	& &   &	&	&   &	&  &  & F10\\
Fit D1			& & (0,\nodata)	&	&	&	&   &	&  &  & \\
~~~~~~~~~~~~~~~~b			& &	& 	7	& 	67.91	&	0 		& 0  &	0 & 0 			& 5.80078 \\
~~~~~~~~~~~~~~~~c			& &   &	10	& 	37.97  	&	0 		& 0  &	0 & 0 			& 3.96172 \\
~~~~~~~~~~~~~~~~d			& &  	& 	10	&	23.42  	&	0.09  	& 0  & 0  & 6.00393 	& 2.19911 \\
\hline\\
HR 8799			&	1.5		& &	&	&	&	&	&   &   & R09\\
Fit D2			& & (25,20)	& 	&	&	&	&	&   &   & \\
~~~~~~~~~~~~~~~~b			& &  	&	5	& 71.2002	 &   0	&	 0 	&	 0 	&	 0 & 0.789567\\
~~~~~~~~~~~~~~~~c			& & 	& 	7	& 41.1437  &   0	&	 0 	&	 0 	&	 0 &  -1.13722\\
~~~~~~~~~~~~~~~~d			& & 	&	7	& 24.4592  &   0	&	 0 	&	 0 	&	 0 &  -3.13463\\
\hline\\
HR 8799 	 		& 	1.5		& &	& & &	& 	&	&	&  R09 \\
Fit D3			& & (20,45)	& 	& & &	& 	&	&	&  \\
~~~~~~~~~~~~~~~~b			& & 	&	5	& 68.3211 &   0	&	 0 	&	 0 	&	 0 &   0.322337 & \\
~~~~~~~~~~~~~~~~c			& & 	&	7	& 40.4333 &	0	&	 0 	&	 0 	&	 0 &  -1.53703 & \\
~~~~~~~~~~~~~~~~d			& & 	&	7	& 24.7391 &	0	&	 0 	&	 0 	&	 0 &  -3.59566 & \\
\hline\\
HR 8799  						&  	1.445	& & 	&	& &	& 	&	&	&  G09 \\
Fit D4			& &  (15.5,11.2)	& 	& & &	& 	&	&	&  \\
~~~~~~~~~~~~~~~~b					& & 	&	8.022	& 68.448 &   0.008	&	 0 	&	 0 	&	 5.38591 & 3.3458 & \\
~~~~~~~~~~~~~~~~c			& & 	&	11.87	& 39.646 &   0.012	&	 0 	&	 0 	&	 6.17550 & 0.698655 & \\
~~~~~~~~~~~~~~~~d			& & 	&	8.891	& 24.181 &   0.075	&	 0 	&	 0 	&	 2.51991 & 2.22704 \\
\hline\\
HR 8799 						& 	1.611	& &	& & &	& 	&	&	&  G09 \\
Fit D5 			&	& (11.4,357.2)	& & &	& 	&	&	&  \\
~~~~~~~~~~~~~~~~b			& & 	&	9.708 & 67.661 & 0.014 &	 0 	&	 0 	& 0.517857 & 2.16072 & \\
~~~~~~~~~~~~~~~~c			& & 	&	7.963 & 31.045 & 0.248 &	 0 	&	 0 	& 4.24831 & 2.77158 & \\
~~~~~~~~~~~~~~~~d			& & 	&	7.397 & 30.777 & 0.267 &	 0 	&	 0 	& 6.08038 & 5.69850 & \\
\enddata
\tablecomments{Planetary parameters used in the dynamical simulations in Section \ref{dyn}. $\it{a}$ and $\it{e}$ are the 
semi-major axis and eccentricity of the planet; in all cases the orbits considered are co-planar ($i$ = 0); $\omega$ is the longitude of periastron; 
$\Omega$ is the longitude of the ascending node; $M$ is the mean anomaly.  The orientation of the orbit is given by angles $I$ and $\Omega$': $I$ is the inclination of the orbital plane with respect to the plane of the sky, and  $\Omega$' is the angle between the North direction and the line of nodes (measured toward the East -- see Fig. 1 in Reidemeister et al. 2009). References are: B06: Butler et al. (2006); G06: Gozdziewski \& Konacki (2006); C09: Couetdic et al. (2009); F10: Fabrycky \& Murray-Clay (2010); B08: Beauge et al. (2008); R09: Reidemeister et al. (2009); G09 Gozdziewski \& Migaszewski (2009). 
}
\end{deluxetable*}

\section{SED Modeling}
\label{sed}

We modeled the excess dust emission using the radiative transfer code developed by Wolf \& Hillenbrand (2003).  For a study on how the SEDs depend on the grain composition we refer to Wolf \& Hillenbrand (2003) and Moro-Mart\'in, Wolf \& Malhotra (2005). For the grain properties, we assumed a silicate composition with optical constants from Weingartner \& Draine (2001), and two assumptions for the particle radius ({\it b}):

\begin{itemize} 
\item  {\it A single grain size of {\it b} = 10 $\mu$m}. This size was chosen  because these grains emit efficiently at 70 $\mu$m. We favor this scenario because it represents large grains in general, where the grains would be located at the blackbody equilibrium distance from the central star (modulo bulk optical properties). Given the lack of spectral features in most debris disks (see e.g. Beichman et al. 2006 and Lawler et al. 2009), it is inferred that small grains with super-thermal behavior are mostly absent, therefore, this single grain size scenario  provides a plausible limiting case for the placement of the grains. 
\item  {\it A particle size distribution following a power law}. The placements of the emitting zones are rather uncertain. In addition to the single grain size scenario above, the other bounding case would be given by assuming a power law down to the blow-out size.  We used $\it{n(b)}$ $\propto$ $\it{b^{-3.5}}$, with  $\it{b_{max}}$ = 10 $\mu$m and $\it{b_{min}}$ = $\it{b_{blow}}$. The blow-out size,  $\it{b_{blow}}$, is the grain radius for which $\beta$, the ratio between the radiation pressure force and the gravitational force, equals 1/2; for spherical grains, $\beta = {3L_* \over 16\pi GM_*c}{<Q_{pr}> \over \rho b} = 5.7\cdot10^{-5} {<Q_{pr}> \over \rho b} {L_* \over L_{\odot}} {M_{\odot} \over M_*} $ (Burns, Lamy \& Soter, 1979), where $\rho$ and $b$ are the density and radius of the grain in cgs units and Q$_{pr}$ is the radiation pressure coefficient, a measure of the fractional amount of energy scattered and/or absorbed by the grain. Q$_{pr}$ is a  function of the physical properties of the grain and the wavelength of the incoming radiation; the value we use, $<Q_{pr}>$,  is an average integrated over the stellar spectrum. 
\end{itemize}

The presence of larger and colder grains is not constrained by the {\it Spitzer} observations, so generally these models lead to lower limits of the dust disk mass. 

We now discuss our assumptions for the disk geometry. We adopted a constant surface density disk ($\Sigma \propto r^0$). The dust disk outer radius, R$_{out}$, cannot be determined from the {\it Spitzer} data alone. Given that the debris disks around HD 128311, HD 202206 and HD 82943 are not spatially resolved at 70 $\mu$m (with a PSF FWHM of 18''), and that the debris disk around $\epsilon$ Eri (of similar spectral type) was well constrained by the {\it Spitzer} data to be $\sim$100 AU (Backman et al. 2009), we considered three possible outer disk radii: R$_{out}$ = 10 AU, 50 AU and 100 AU. To explore different belt widths, the inner disk radius, R$_{in}$, is allowed to vary from the sublimation radius (where T$_{sub}$ = 1550 K) to R$_{out}$. 

We further assumed that the dust disk is optically thin (which is supported by the low fractional luminosities observed for HD 128311, HD 202206 and HD 82943, with L$_{dust}$/L$_{star}$ $<$  10$^{-4}$ -- see  Table \ref{multidisks}). Under this scenario, only scattering, absorption and reemission of stellar radiation by dust grains were taken into account (neglecting scattering and dust heating from the dust infrared radiation). With the parameters described above, we calculated the dust disk emission at 200 wavelengths from 3 $\mu$m to 600 $\mu$m.  

\begin{figure}
\epsscale{0.9}
\plotone{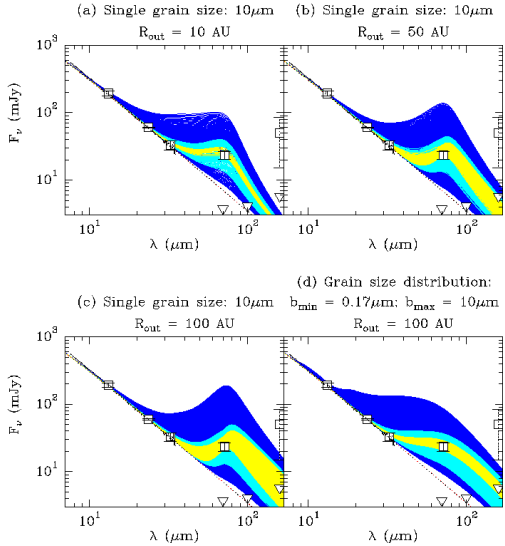}
\caption{
Observed and modeled SEDs for HD~128311. The {\it dotted} line is the stellar photosphere. The {\it Spitzer} photometric points ({\it MIPS} and synthetic photometry from {\it IRS}) are represented by {\it squares} with 1-$\sigma$ error bars. The continuous lines correspond to the modeled SEDs and include the emission from the photosphere and from a disk of dust composed of particles with optical properties typical of astronomical silicates. Panels (a), (b) and (c) correspond to 10 $\mu$m size grains. Panel (d) corresponds to a grain size distribution given by $\it{n(b)}$ $\propto$ $\it{b^{-3.5}}$ and a maximum grain radius of $\it{b_{max}}$ = 10 $\mu$m; regarding the minimum grain radius, we adopt the arbitrary value of $\it{b_{min}}$ = 0.17 $\mu$m because the low luminosity of HD 128311 results in the absence of a blow-out size.  The dust disk has a mass M$_{dust}$ and extends from R$_{in}$ to R$_{out}$ with a constant surface density. R$_{in}$ and M$_{dust}$ are the  free parameters; R$_{out}$ is kept fixed at 10 AU, 50 AU and 100 AU (indicated at the top of each panel).  The models in  $\it{yellow}$ are those with a $\chi$$^{2}$ probability P($\chi$$^{2}$ $\mid$ $\nu$) $<$ 0.683;  $\it{light~blue}$ for P($\chi$$^{2}$ $\mid$ $\nu$) $>$ 0.683, i.e. models that can be excluded with 1-$\sigma$ certainty; and $\it{dark~blue}$ for P($\chi$$^{2}$ $\mid$ $\nu$) $>$ 0.9973, i.e. models that are excluded with 3-$\sigma$ certainty. The {\it triangles} correspond to the 5-$\sigma$/1hr sensitivity limits for Herschel/PACS: 3.75 mJy (60 $\mu$m--85 $\mu$m), 4.1 mJy (85 $\mu$m--130 $\mu$m) and 5.75 mJy (130 $\mu$m--210 $\mu$m). 
 }
\label{sed_hd128311}
\end{figure}

\begin{figure}
\epsscale{0.83}
\plotone{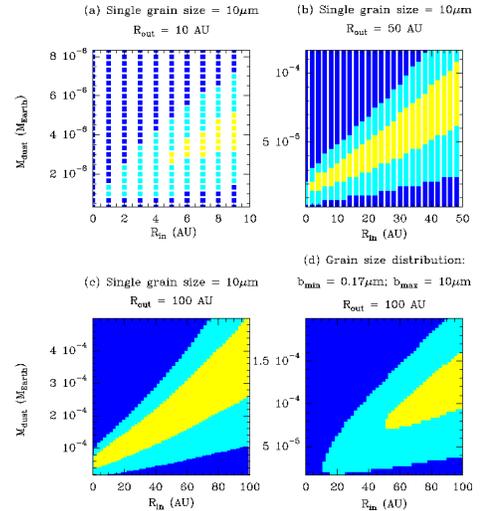}
\caption{Parameter space of the modeled SEDs in Figure \ref{sed_hd128311}. Each point of these two-dimensional grids represents 
a modeled SED from Figure \ref{sed_hd128311}, where R$_{in}$ and M$_{dust}$ are the two free parameters.  
The models in  $\it{yellow}$ are those with a $\chi$$^{2}$ probability P($\chi$$^{2}$ $\mid$ $\nu$) $<$ 0.683; 
$\it{light~blue}$ for P($\chi$$^{2}$ $\mid$ $\nu$) $>$ 0.683, i.e. models that can be excluded with 1-$\sigma$ certainty; and $\it{dark~blue}$ for P($\chi$$^{2}$ $\mid$ $\nu$) $>$ 0.9973, i.e. models that are excluded with 3-$\sigma$ certainty. 
} 
\label{chi_hd128311}
\end{figure}

\subsection{HD 128311}
\label{sed_HD128311}
Beichman et al. (2006) modeled the {\it IRS} and {\it MIPS} 70 $\mu$m data with a single population of 10 $\mu$m amorphous silicate grains located at R$_{dust}$ $>$ 15 AU (T$_{dust}$ $<$ 50 K), while the modeling of the {\it MIPS} 24 $\mu$m and 70 $\mu$m data in Trilling et al. (2008) implied R$_{dust}$ $>$  5.1 AU (T$_{dust}$ $<$ 106 K), with a fractional luminosity L$_{dust}$/L$_{star}$ = 1.3--2.7$\cdot$10$^{-5}$.

To model the dust emission, the stellar contribution needs to be subtracted from the observed SED. Because the slope of the {\it IRS} spectra of HD 128311 indicates there is no dust excess at  $\lambda$$<$ 34 $\mu$m,  we pin the photosphere to the MIPS 24 $\mu$m flux; for simplicity, the stellar emission was modeled as  a blackbody with T$_{eff}$ = 4965 K,  L$_*$ = 0.24 \lsol~ and a distance of 16.57 pc.  

For a stellar mass of M$_{*}$ = 0.84 \msol,  a grain density $\rho$ = 2.5 g cm$^{-3}$,  and adopting the optical constants of astronomical silicates, we find that the maximum $\beta$-value is 0.26, i.e. there is no blow-out size because  $\beta<$ 0.5. 

Figure \ref{sed_hd128311}  shows the SED models compared to the SED observed by {\it Spitzer}. 
We calculated the synthetic photometry of each SED model using the {\it MIPS} and {\it IRS} filter profiles from the {\it Spitzer} Users Manual, with effective wavelengths at 13.2 $\mu$m, 23.68 $\mu$m, 32.5 $\mu$m, 71.42 $\mu$m and 155.9 $\mu$m. Then, for each model we calculated  $\chi$$^2$ and the $\chi$$^2$ probability, where the fit is done to the 5 photometric points in Table \ref{spitphotom} with two free parameters (R$_{min}$ and M$_{dust}$).  [Because the uncertainty in the 160 $\mu$m flux is large, even though we are including it in the SED fit, in practice its contribution is negligible]. The colors indicate the goodness of the fit.
Figure \ref{chi_hd128311} shows the parameter space of the SED models in Figure \ref{sed_hd128311}. The {\it yellow} area corresponds to combinations of R$_{in}$ and M$_{dust}$ that lead to a $\chi$$^2$ probability of P($\chi$$^{2}$ $\mid$ $\nu$) $<$ 0.683, $\it{light~blue}$ is for 0.683 $<$ P($\chi$$^{2}$ $\mid$ $\nu$) $<$ 0.9973, and $\it{dark~blue}$ if for P($\chi$$^{2}$ $\mid$ $\nu$) $>$ 0.9973.  The modeling of the SED is degenerate.  For the single grain size models,  valid SED fits can be obtained for: R$_{out}$ = 10 AU and R$_{in}$ $\ge$ 5 AU;  R$_{out}$ = 50 AU and R$_{in}$ $\ge$ 2 AU; and R$_{out}$ = 100 AU and R$_{in}$ $\ge$ 2 AU.
In all three cases, narrow belts with widths of 10\% the disk radius could also fit the observed SEDs. Here, we would like to emphasize the lower limits to R$_{in}$ because of two reasons: (1) For HD 128311, HD 202206 and HD 82943 the dust is located outside the orbits of the planets and, therefore, it is at the disk inner edge where the gravitational effects of the planets are stronger (i.e. it is at that location that there is a closer correspondence between the SED models and the dynamical models). (2) The narrowest belts might only be justified in the presence of additional planetary perturbers for which we have no evidence so far.

As we mentioned before, we favor the single grain  size models above because they represent large grains in general, where the grains would be located at the blackbody equilibrium distance from the central star. However, for completeness and to explore the other limiting case we have also considered the case of a grain size distribution.  
For these models, valid fits can be obtained for R$_{out}$ = 100 AU and R$_{in}$ $\ge$ 52 AU. A small disk is ruled out for this grain size distribution.  We can exclude the presence of a significant population of small grains unless the disk is large, 100 AU,  in which case there would be a depletion of small grains inside 52 AU.

\begin{figure}
\epsscale{0.88}
\plotone{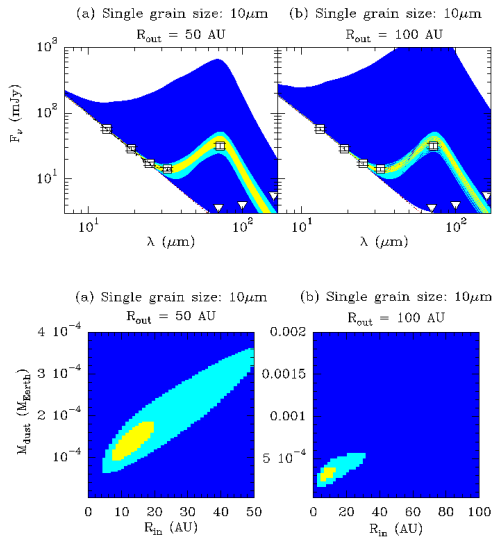}
\caption{Same as Figures \ref{sed_hd128311}  and \ref{chi_hd128311} but for HD~202206. 
Of the range of models explored, only the two sets of models shown in the panels could fit the observations.
}
\label{sed_chi_hd202206}
\end{figure}

\subsection{HD 202206}
\label{sed_HD202206}
The slope of the {\it IRS} spectrum of HD 202206 indicates there is no dust excess at  $\lambda$$<$ 25 $\mu$m; therefore, we used the IRS synthetic photometry at 15--19 $\mu$m to determine the photospheric emission level. The stellar emission was modeled as  a blackbody with T$_{eff}$ =  5764 K,  L$_*$ = 0.92 \lsol~ and a distance of 46.3 pc. 

For a stellar mass of M$_{*}$ = 1.044 \msol,  a grain density $\rho$ = 2.5 g cm$^{-3}$  and adopting the optical constants of astronomical silicates, the blow-out size  is $\it{b_{blow}}$ = 0.5 $\mu$m. 

For the SED modeling, we followed the same scheme as described above for HD 128311.  Figure \ref{sed_chi_hd202206} (top)  shows some of the SED models  computed; over-plotted are the  {\it Spitzer} observations.  For each model, we calculated the synthetic photometry  using the {\it MIPS} and adopted {\it IRS} filter profiles at effective wavelengths 13.2 $\mu$m, 19 $\mu$m, 25 $\mu$m, 32.5 $\mu$m and  71.42 $\mu$m. We then calculated  $\chi$$^2$ and the $\chi$$^2$ probability distribution, where the fit is done to the observed 5 photometric points in Table \ref{spitphotom} with two free parameters (R$_{min}$ and M$_{dust}$). The colors indicate the goodness of the fit. Of the wide range of models explored, only the two sets of models shown can fit the observations. Figure \ref{sed_chi_hd202206} (bottom) shows the parameter space  (R$_{min}$ and M$_{dust}$) of the models in the top two panels. The SED modeling is degenerate, but compared to HD 128311, the case of HD 202206 is better  constrained because of the detection of an excess starting near 25 $\mu$m.  For the models that assume a single grain size, the best fits are given by: R$_{out}$ = 50 AU and 10 AU $\le$ R$_{in}$ $\le$ 20 AU; and R$_{out}$ = 100 AU and 6AU $\le$ R$_{in}$ $\le$ 10 AU. For this case we can  rule out the presence of  a small 10 AU disk. 
We  can also rule out the models that assume a distribution of grain sizes with a power-law index of -3.5,  $\it{b_{min}}$ = $\it{b_{blow}}$ and $\it{b_{max}}$ = 10 $\mu$m. The latter might not surprising because collisional disk models by several authors (e.g. Krivov et al. 2006; Thebault \& Augereau 2007; Mueller et al. 2010, Krivov  2010) conclude that the size distribution shows substantial deviations from a power-law near the blowout size regime, suggesting that $b_{min}$ would need to be two or three times larger than $b_{blow}$, and that the power-law index may be smaller than 3.5, to mimic a secondary maximum seen in the size distribution at hundreds of microns. 

\begin{figure}
\epsscale{0.88}
\plotone{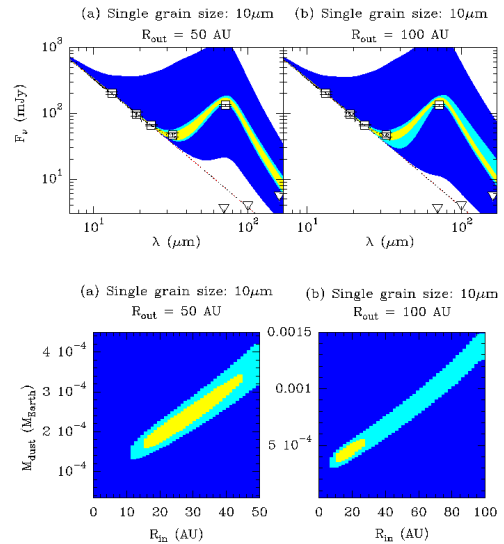}
\caption{Same as Figures \ref{sed_hd128311}  and \ref{chi_hd128311} but for HD 82943. Of the range of models explored, only the two sets of models shown in the panels could fit the observations.
}
\label{sed_chi_hd82943}
\end{figure}

\subsection{HD 82943}
\label{sed_HD82943}

Because the {\it IRS} spectrum of HD 82943 indicates there is no dust excess at  $\lambda$$<$ 22 $\mu$m, we used the {\it IRS} synthetic photometry flux calculated between 15--19 $\mu$m to pin down the photosphere.  For simplicity, the stellar emission was modeled as  a blackbody with T$_{eff}$ = 5989 K,  L$_*$ = 1.25 \lsol, and a distance of 27.46 pc. 

For a stellar mass of M$_{*}$ = 1.15 \msol,  a grain density $\rho$ = 2.5 g cm$^{-3}$  and adopting the optical constants of astronomical silicates, the blow-out size  is $\it{b_{blow}}$ = 0.6 $\mu$m. 

Figure \ref{sed_chi_hd82943}  (top) shows  the SED models  over-plotted on the  {\it Spitzer} observations.  For each model, we calculated the synthetic photometry  using the {\it MIPS} and adopted {\it IRS} filter profiles at effective wavelengths of 13.2 $\mu$m, 19 $\mu$m, 23.68 $\mu$m, 32.5 $\mu$m and  71.42 $\mu$m. We then calculated  $\chi$$^2$ and the $\chi$$^2$ probability distribution, where the fit is done to the observed 5 photometric points in Table \ref{spitphotom} with two free parameters (R$_{min}$, M$_{dust}$).  The colors indicate the goodness of the fit. Figure \ref{sed_chi_hd82943} (bottom) shows the parameter space of the models in the top panels. 

For the models that assume a single grain radius of 10 $\mu$m, the best SED fits are given by: R$_{out}$ = 50 AU and 16 AU $\le$ R$_{in}$ $\le$ 44 AU; and R$_{out}$ = 100 AU and 12 AU $\le$ R$_{in}$ $\le$ 26 AU. For this case we can  rule out the presence of  a small 10 AU disk.  As for HD 202206, we can also rule out models that assume a grain size distribution with a power-law index of -3.5,  $\it{b_{min}}$ = $\it{b_{blow}}$ and $\it{b_{max}}$ = 10 $\mu$m. 

\subsection{HR 8799}
\label{sed_HR8799}

The {\it Spitzer} observations of HR 8799 and the analysis of its SED and spatially resolved images were presented in Su et al. (2009). Here we summarize the main results. 
The stellar emission is best fitted by a Kurucz model with  T$_{eff}$ =  7500 K,  R$_*$ = 1.4 R$_{\odot}$, log($g$) = 4.5 and sub-solar abundances, with a resulting L$_*$ = 5.7 \lsol~(consistent with a young age). The stellar mass is M$_*$ = 1.5 \msol. As in Sections \ref{sed_HD128311} and \ref{sed_HD202206}, Su et al. (2009) assumed spherical grains with optical properties characteristic of astronomical silicates, a bulk density of 2.5 g cm$^{-3}$, and a size distribution following $\it{n(b)}$ $\propto$ $\it{b^{-3.5}}$;  the blow-out size for HR 8799 is  2 $\mu$m. 
Su et al. (2009) found that the stellar-subtracted images and SED (after applying color-corrections) can be best fit by the following multi-component disk model:
\begin{itemize}

\item {\it Unresolved warm disk with a characteristic temperature of $\sim$150 K}. \\
This component was modeled with a flat surface density, $\Sigma$ $\sim$ $r^{0}$. The maximum inner radius is R$_{in}$ = 6 AU (based on temperature arguments) and the outer radius is R$_{out}$  = 15 AU. The grain sizes range from 1 to 4.5 $\mu$m (with the spectral shape between 10 and 20 $\mu$m requiring the presence of $\mu$m-sized grains and the absence of sub-micron grains). If we were to assume all the grains are bound ($>$ 2 $\mu$m), the outer radius of this warm component would be R$_{out}$ $\sim$ 10 AU (instead of 15 AU). The estimated  dust mass is 1.1$\cdot$10$^{-6}$ M$_{\oplus}$. This is a lower limit because it is not possible to constrain the mass contributed by larger grains as there are no long wavelength detections of this warm component. This component dominates the emission at {\it IRS} wavelengths and the unresolved emission at 24 $\mu$m (it is too warm to contribute to the unresolved emission at 70 $\mu$m). 
The upper limit to the dust mass $<$ 6 AU is 2.3$\cdot$10$^{-7}$ M$_{\oplus}$ (assuming 10 $\mu$m grains). 
 
\item {\it Unresolved cold disk with a characteristic temperature of $\sim$45 K}.\\
This component was also modeled with a flat surface density, $\Sigma$ $\sim$ $r^{0}$. The inner radius, R$_{in}$ = 90 AU, is not constrained from imaging but from the SED: derived from the characteristic temperature of the cold component assuming blackbody grains 10 $\mu$m in size. The sharp inflection between the {\it IRS} and the {\it MIPS}-SED spectra indicates that there is a sharp inner edge of the cold dust component (i.e. there is very little dust located between the warm and the cold dust components). The outer radius is set to R$_{out}$  = 300 AU, but is less well constrained because of the lack of spatially resolved sub-mm observations; the value of 300 AU is chosen so that the $\mu$m-sized grains in the extended halo (discussed next), assumed to originate from this cold component, are not too warm as to make the emission at 25--35 $\mu$m inconsistent with the observations.
The grain sizes range from 10 to 1000 $\mu$m. Large grains are assumed to exist because this is the location of the collisional cascade that populates the extended halo discussed below.  The estimated dust mass of this component is 0.12 M$_{\oplus}$. This component is needed to fit the SED $>$ 30 $\mu$m. It dominates the unresolved emission at 70 $\mu$m and constitutes $<$9\% of the total (unresolved+resolved) emission at 24 $\mu$m. This component also accounts for the unresolved sub-mm emission in Williams \& Andrews (2006).

\item {\it Extended halo}.\\ 
At 24 $\mu$m the cold component discussed above would be barely resolved, and a 70 $\mu$m it would be unresolved, therefore, it is inferred that most of the resolved 24 $\mu$m and 70 $\mu$m emission comes from an extended halo component. It is modeled with a surface density, $\Sigma$ $\sim$ $r^{-1}$, characteristic of an unbound disk.  The limited spatial resolution at 70 $\mu$m does not allow determination of the inner radius; it is placed at R$_{in}$ = 300 AU because it is assumed the grains in the halo originate in a collisional cascade in the the unresolved cold disk. R$_{out}$  is set to 1000 AU because the disk can be traced even beyond this distance (the best-fit outer radius ranges from 900 AU to 1800 AU).  The grain sizes range from 1 to 10 $\mu$m (grains larger than the blow-out size are included to account for the effects of porosity on the response of the grains to radiation pressure). Assuming a grain size distribution following $\it{n(b)}$ $\propto$ $\it{b^{-3.5}}$ and a maximum grain size of 10 $\mu$m, the ratio of the 70 and 24 $\mu$m fluxes of the extended component is consistent with a  minimum grain size of $\lesssim$ 2 $\mu$m (similar to the blow-out size for HR 8799). 
The estimated dust mass of this component is 1.9$\cdot$10$^{-2}$ M$_{\oplus}$. 
\end{itemize}

In Section \ref{loc_HR8799} we discuss the above results from Su et al. (2009)\footnote {Reidemeister et al. (2009) also carried out an analysis of the SED of HR8799. They assume a flat surface density distribution ($\Sigma$ $\sim$ $r^{0}$), astronomical silicate composition, and particle size distribution of $\it{n(b)}$ $\propto$ $\it{b^{-3.5}}$ with $b_{min} = 5 \mu$m and $b_{max} = 1000 \mu$m. They inferred the presence of a warm component with R$_{out}$ = 10--15 AU, and a cold component with R$_{in}$ = 75--120 AU and R$_{out}$ = 125--170 AU. We will take this result with caution because: (1) this fit does not include the deep {\it IRS} and spatially resolved {\it MIPS} observations (published after Reidemeister et al. 2009 paper); (2) they noted that the calibration of the (shallower) {\it IRS} data used is uncertain; and (3) we noted that their color-corrections for the {\it IRAS} observations were done for a characteristic temperature of 5000 K, instead of the much smaller dust temperature (that would lead to smaller correction factors).}
in the context of the dynamical simulations in Section \ref{dyn_HR8799}; these simulations consider the five possible planetary configurations listed in Table \ref{planetparam} and described in Section \ref{obs_HR8799}.
\\
\\
\section{Dynamical Modeling}
\label{dyn}
\subsection{HD 128311}
\label{dyn_HD128311}

\begin{figure}
\epsscale{1.1}
\plottwo{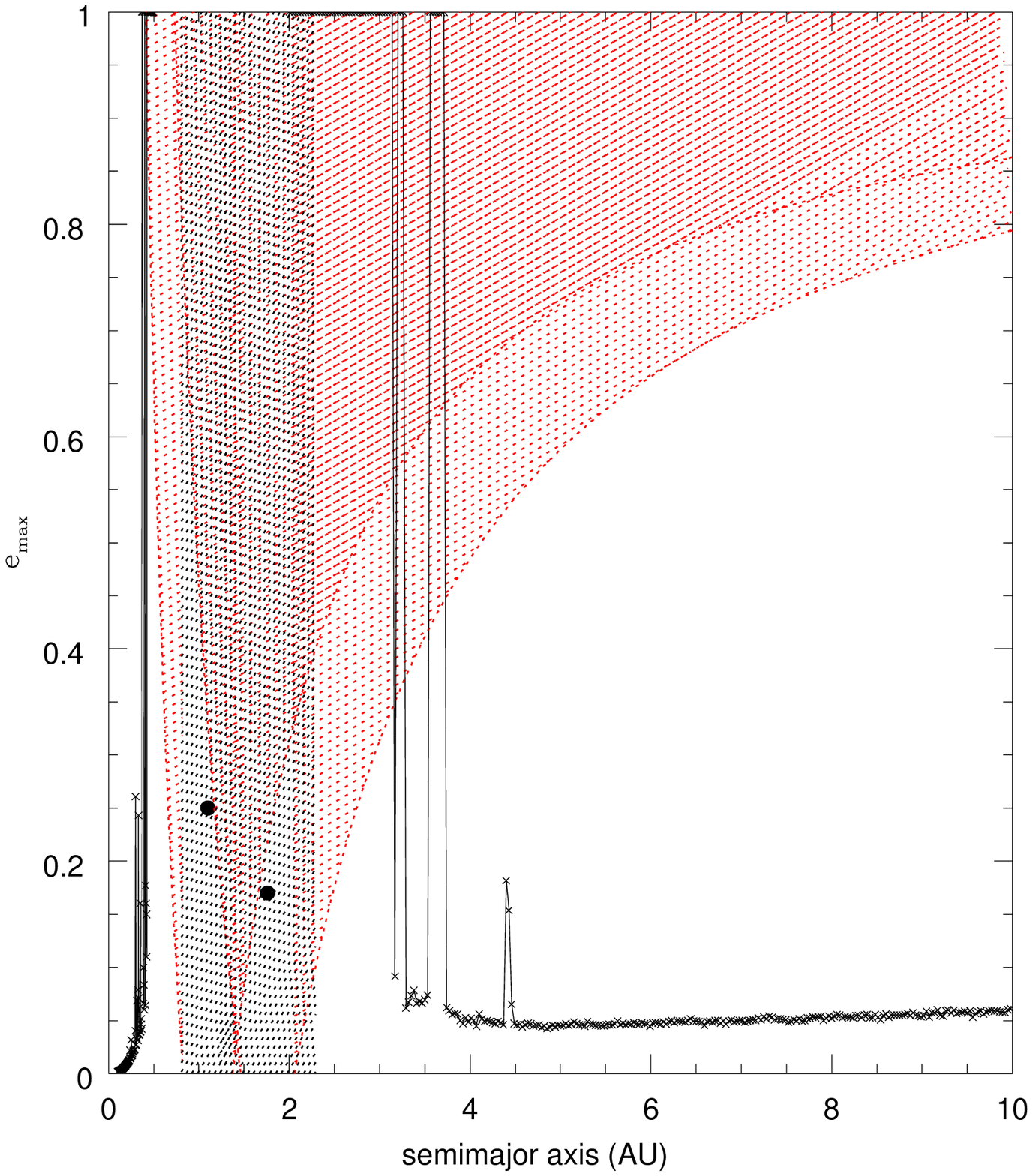}{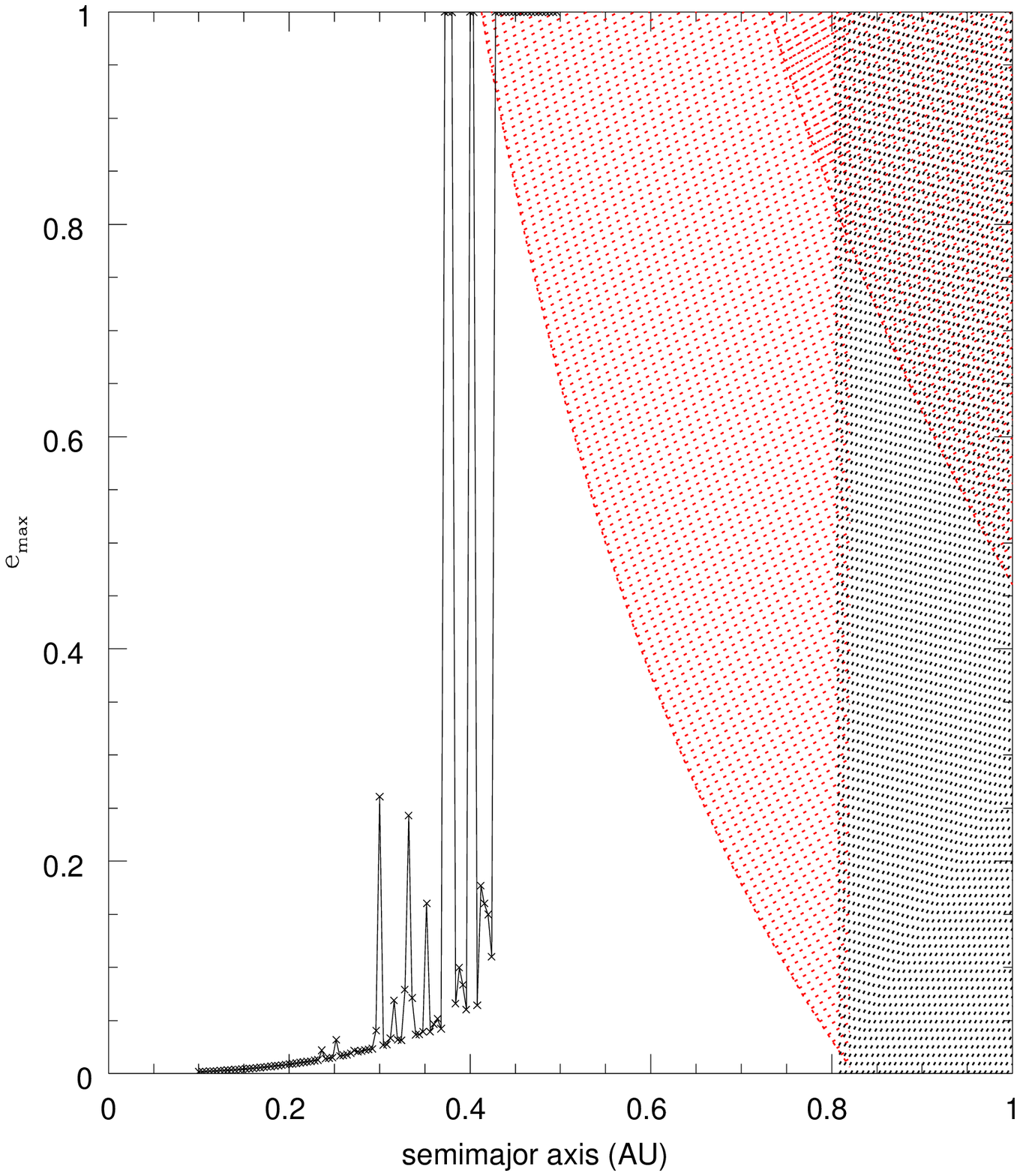}
\caption{Allowed parameter space for the planetesimals' orbital elements. {\it Shaded area}: regions where test particle orbits are unstable due to planet-crossing (red) or overlapping first order mean motion resonances (grey).  The large black dots represent the two planets. The connected crosses are the maximum eccentricity attained by test
particles on initially circular orbits, resulting from a numerical  integration lasting 1 Myr.  A zoom in to the region interior to the planets is shown in the right panel.  The zone inside $\sim$0.3 AU and the zone beyond $\sim$ 4 AU appears stable for potential locations of planetesimal belts. 
}
\label{sim_hd128311_renu}
\end{figure}

\begin{figure}
\begin{center}
\includegraphics[scale=0.33,angle=0]{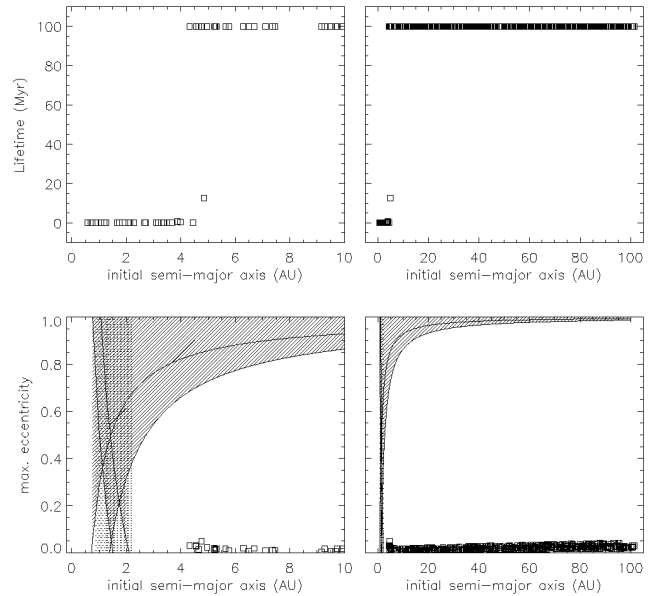}
\end{center}
\caption{Results from the dynamical simulation of 500 test particles in the HD 128311 planetary system (Fit A2). A zoom in to the inner system is shown in the left panels. The numerical integration lasted 100 Myr. {\it Top}: test particle's lifetimes. {\it Bottom}: allowed parameter space for the planetesimals' orbital elements, where the shaded areas indicate regions where test particle's orbits are unstable due to planet-crossing ({\it striped area}) or overlapping first order mean motion resonances ({\it dotted area}); the squared symbols show the maximum eccentricity attained by test particles on initially circular orbits.  
} 
\label{HD128311_dyn_A2}
\end{figure}

Using numerical simulations we have studied the effect of short and long-term planetary perturbations on the orbital stability of the dust-producing planetesimals. 
Notice that with the orbital elements adopted in this paper for HD 128311 (Models A1 and A2 in Table \ref{planetparam}), the planets are just barely outside their own planet-crossing zones. Because of the proximity of the planets' orbits to each other and to the 2:1 resonance, the analytical theory for the secular perturbations cannot be applied (as it was done for HD 38529 in Moro-Mart\'in et al. 2007b), and therefore we study the stability of the orbits using numerical integrations.   Figure \ref{sim_hd128311_renu} corresponds to Fit A1 and shows the maximum eccentricity attained by test particles on initially circular orbits, resulting from gravitational perturbations by the two planets for an integration time of 1 Myr. We can exclude the presence of planetesimals in the regions shaded with {\it black} dots because those orbits are chaotic due to overlapping first-order mean motion resonances (Wisdom 1980); similarly, the regions shaded with {\it red} dots are not expected to have stable planetesimals because they are planet-crossing eccentric orbits. 
The stability of the potential dust-producing planetesimal belt was also studied for the orbital solution in Fit A2  (Figure \ref{HD128311_dyn_A2}).  For this simulation, we used 500 particles uniformly spaced between 0.5 AU and 100 AU, on initially circular orbits co-planar with the planets, with angular elements chosen randomly between 0 and 2$\pi$. Particles were removed if they approached the star closer than 0.5 AU, or approached a planet closer than its Hill radius. 
The orbits were integrated for 
$\sim$ 100 Myr 
using the multiple time step symplectic method skeel-SyMBA (Duncan, Levison \& Lee 1998).  The evolution of the planets' semi-major axes and eccentricities  is shown in Figure \ref{ploe}. 
The potential locations of planetesimal belts are the regions where the orbits are stable and where the maximum eccentricity is low ($\lesssim$0.3), ensuring long planetesimals' lifetimes. Figures \ref{sim_hd128311_renu} and \ref{HD128311_dyn_A2} show that there are two such regions: 
(a) interior to $\sim$ 0.3 AU and (b) exterior to $\sim$ 4 AU (for $a >$ 4 AU the maximum eccentricity always remains below 0.05). 

\subsection{HD 202206}

\begin{figure}
\begin{center}
\includegraphics[scale=0.33,angle=0]{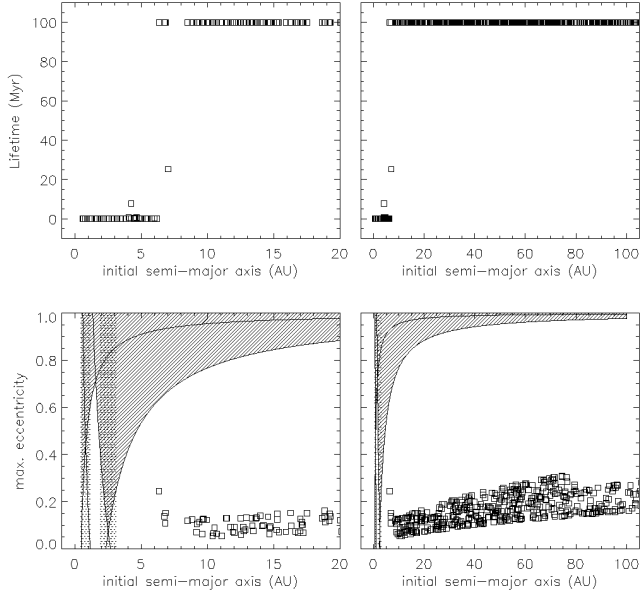}
\end{center}
\caption{Same as Figure \ref{HD128311_dyn_A2} for 500 test particles in the HD 202206 planetary system (Fit B1); the numerical integration lasted 100 Myr.   
} 
\label{HD202206_dyn_B1}
\end{figure}

The stability of the potential dust-producing planetesimals was studied using numerical simulations of test particles with the planetary parameters listed in Table \ref{planetparam} (Fit B1).   
The simulations were done for 500 particles uniformly spaced between 0.5 AU and 100 AU, with an integration time of 100 Myr. 
The evolution of the planets' semi-major axes and eccentricities during that time span is shown in Figure \ref{ploe}. 
The results shown in Figure \ref{HD202206_dyn_B1} indicate that the test particle's orbits are stable beyond $\sim$ 6 AU, and that their maximum eccentricity remains $\lesssim$ 0.3 for all semi-major axis; this indicates that the dust-producing planetesimals are located beyond 6 AU. \\
\\
\\
\\
\subsection{HD 82943}

\begin{figure}
\begin{center}
\includegraphics[scale=0.33,angle=0]{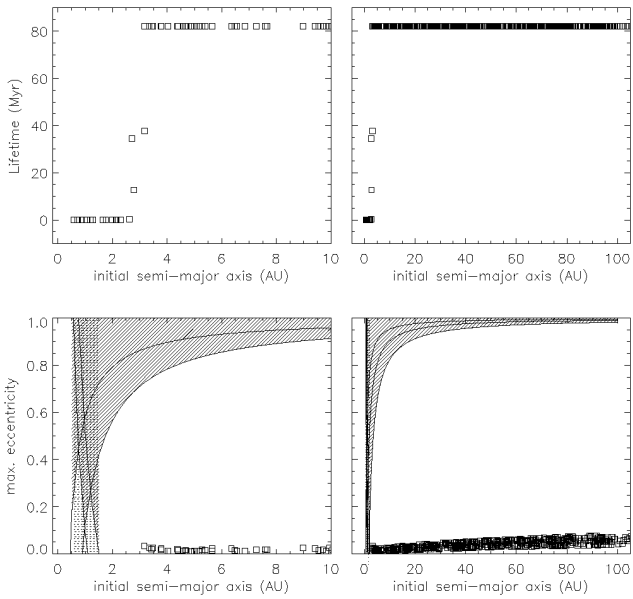}
\end{center}
\caption{Same as Figure \ref{HD128311_dyn_A2} for 500 test particles in the HD 82943 planetary system (Fit C1); the numerical integration lasted $\sim$ 82 Myr.  
} 
\label{HD82943_dyn_C1}
\end{figure}

\begin{figure}
\begin{center}
\includegraphics[scale=0.33,angle=0]{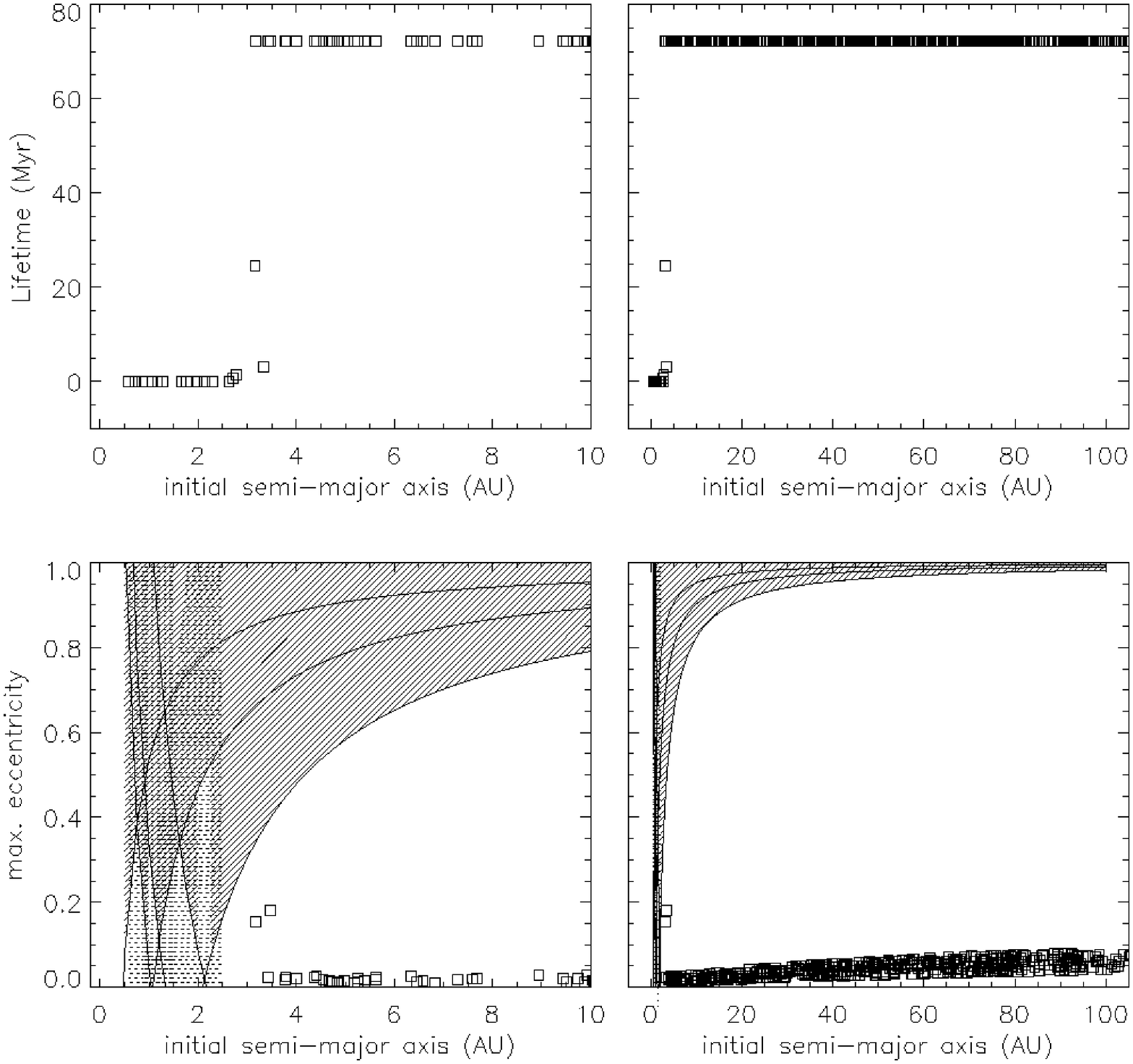}
\end{center}
\caption{Same as Figure \ref{HD128311_dyn_A2}  but for Fit C2 of HD 82943; the numerical integration lasted $\sim$ 72 Myr.  
} 
\label{HD82943_dyn_C3}
\end{figure}

\begin{figure}
\begin{center}
\includegraphics[scale=0.33,angle=0]{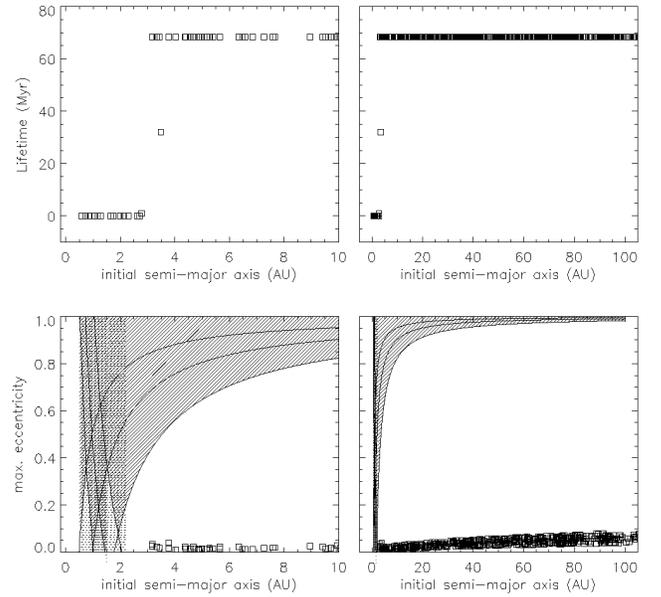}
\end{center}
\caption{Same as Figure \ref{HD128311_dyn_A2}  but for Fit C3 of HD 82943); the numerical integration lasted $\sim$ 68 Myr.  
} 
\label{HD82943_dyn_C4}
\end{figure}

We have studied numerically the stability of the potential dust-producing planetesimals in the three planetary configurations discussed in Section \ref{obs_HD82943} (listed in Table \ref{planetparam}).   
The simulations were done for 500 particles uniformly spaced between 0.5 AU and 100 AU, with an integration time of $\gtrsim$ 70 Myr. 
The evolution of the planets' semi-major axes and eccentricities is shown in Figure \ref{ploe}. 
The results in Figures \ref{HD82943_dyn_C1}--\ref{HD82943_dyn_C4} indicate that, in the three planetary configurations considered, the test particle orbits are stable beyond $\sim$ 3 AU, with maximum eccentricities  always  $<$ 0.1. Long-lived, dust-producing planetesimals could therefore be located anywhere beyond 3 AU.

\label{dyn_HD202206}

\label{dyn_HD82943}

\subsection{HR 8799}
\label{dyn_HR8799}

\begin{figure}
\begin{center}
\includegraphics[scale=0.25]{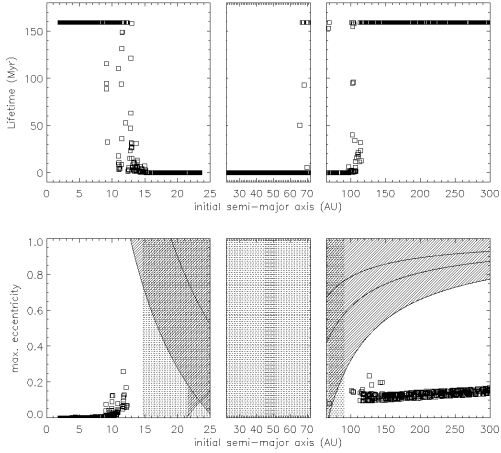}
\end{center}
\caption{Same as Figure \ref{HD128311_dyn_A2} for 1500 test particles in the HR 8799 planetary system (Fit D1); the numerical integration lasted 160 Myr.  The very slight rising trend of the maximum eccentricity at large semi-major axis seen in Figures \ref{HR8799_dyn_D1},  \ref{HR8799_dyn_D2},  \ref{HR8799_dyn_D3},  \ref{HR8799_dyn_D4} and \ref{HR8799_dyn_D5} is a numerical artifact that arises because the test particle simulations for HR8799 were carried out by skeel-SyMBA which gives orbital elements in stellar-centric osculating elements; this produces a spurious eccentricity in nearly circular orbits because the star is wobbling, and its wobble velocity becomes a larger and larger fraction of the orbital velocity of test particles at larger and larger distances. (This effect has been corrected in Figures 8--12, where the osculating keplerian elements of the test particles are calculate in the barycentric frame.)
}
\label{HR8799_dyn_D1}
\end{figure}

\begin{figure}
\begin{center}
\includegraphics[scale=0.30,angle=0]{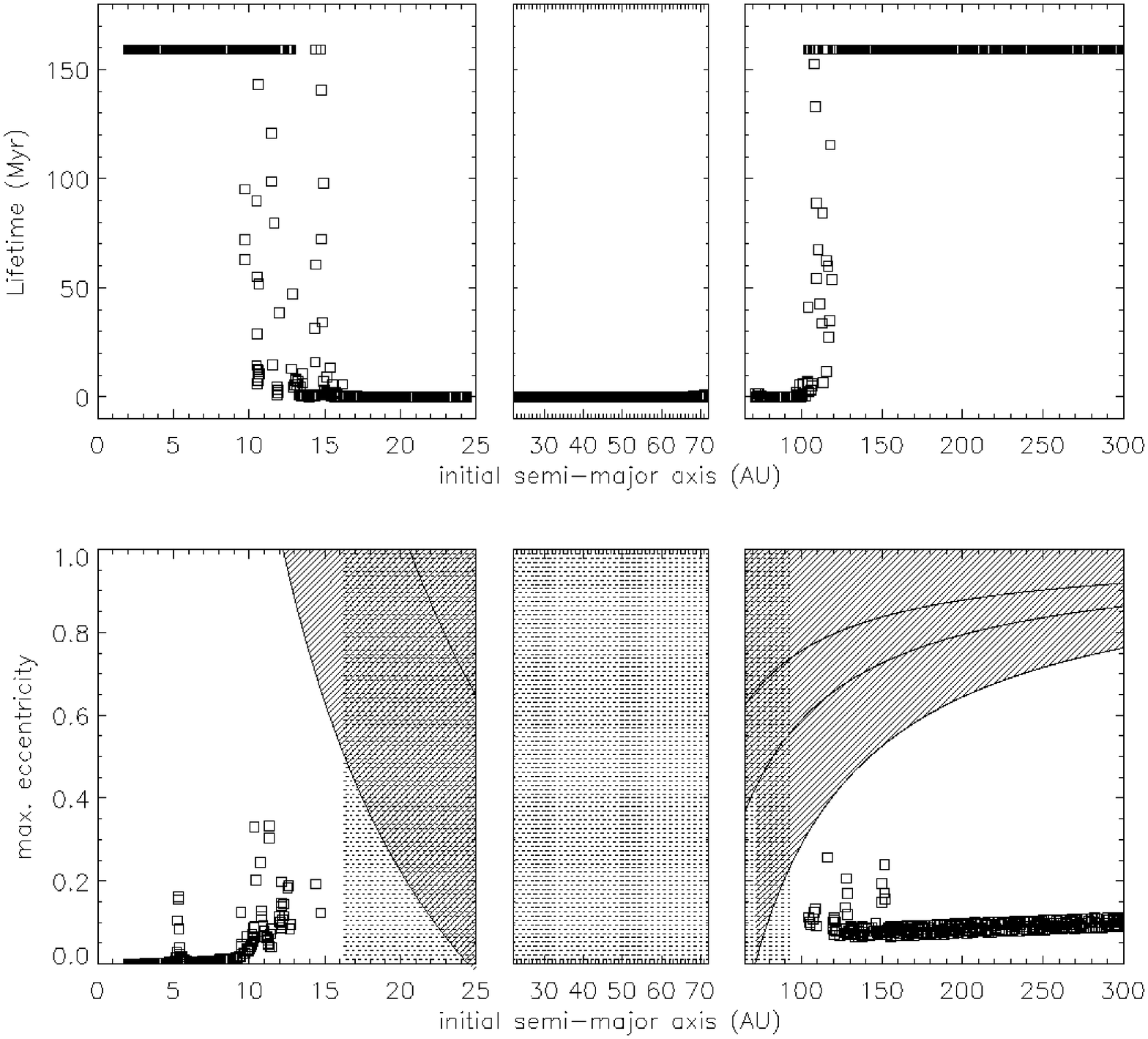}
\end{center}
\caption{Same as Figure \ref{HR8799_dyn_D1} but for Fit D2 of HR 8799.
} 
\label{HR8799_dyn_D2}
\end{figure}

\begin{figure}
\begin{center}
\includegraphics[scale=0.30,angle=0]{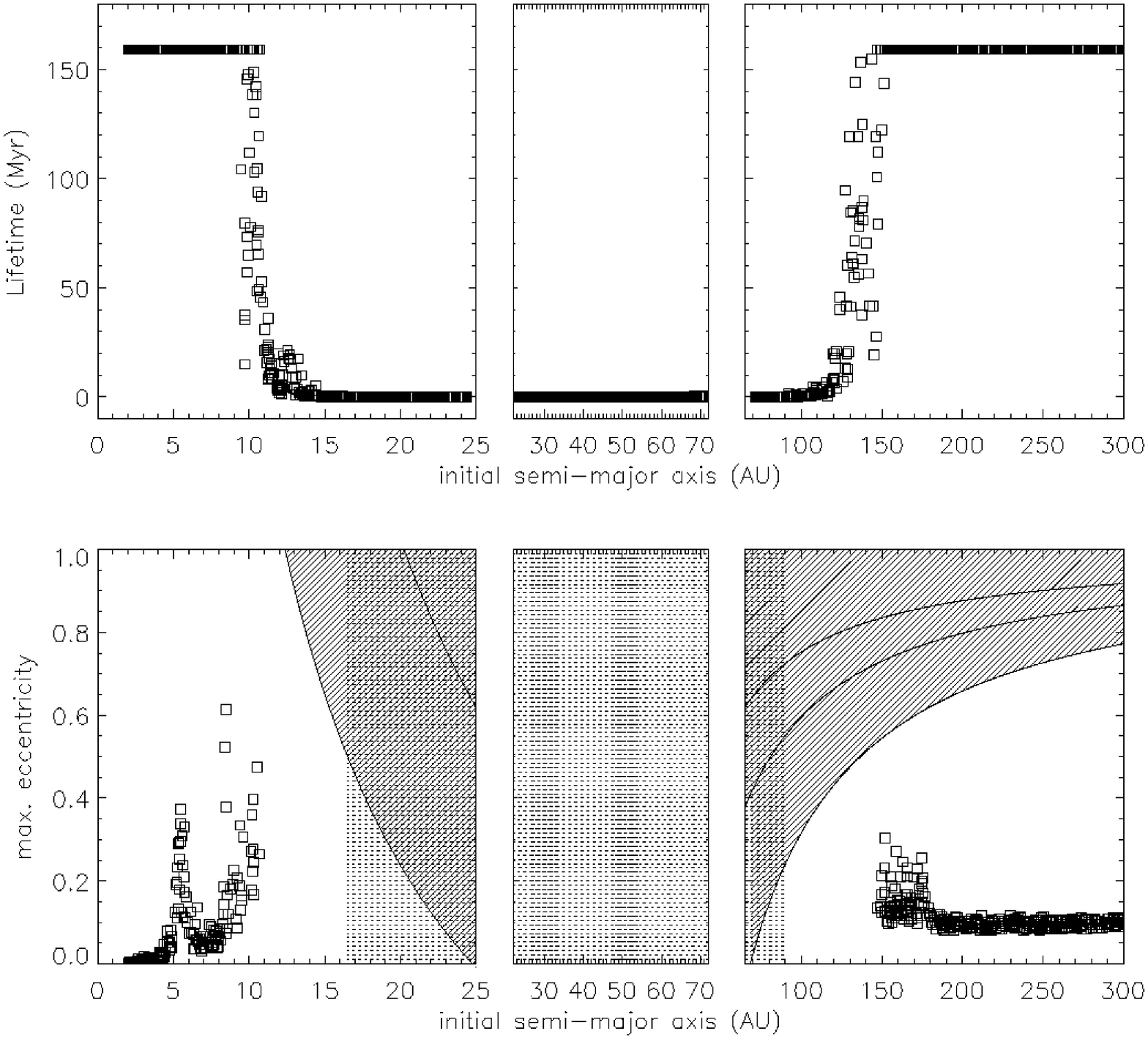}
\end{center}
\caption{Same as Figure \ref{HR8799_dyn_D1} but for Fit D3 of HR 8799.
} 
\label{HR8799_dyn_D3}
\end{figure}

\begin{figure}
\begin{center}
\includegraphics[scale=0.30,angle=0]{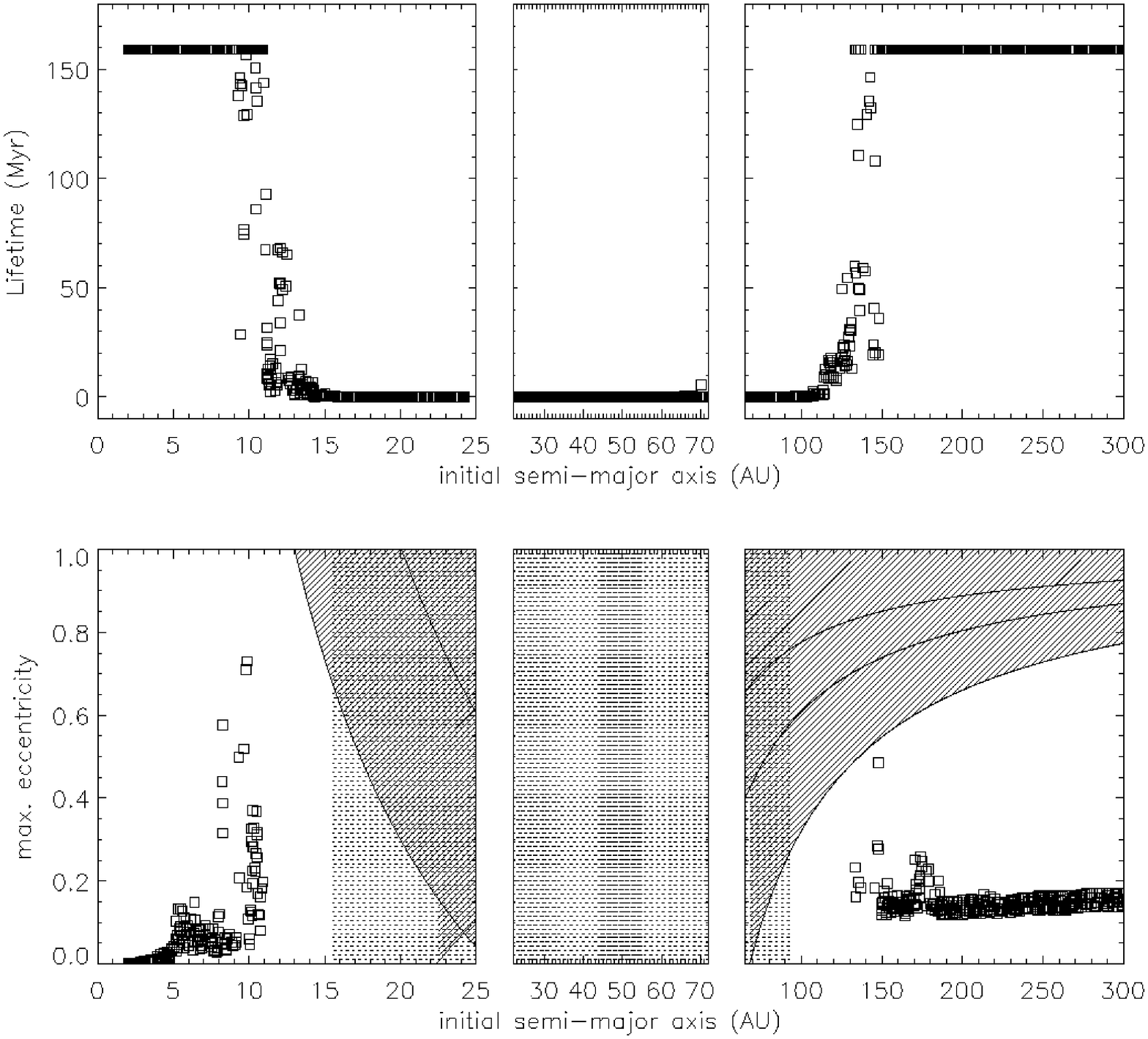}
\end{center}
\caption{Same as Figure \ref{HR8799_dyn_D1} but for Fit D4 of HR 8799.
} 
\label{HR8799_dyn_D4}
\end{figure}

\begin{figure}
\begin{center}
\includegraphics[scale=0.30,angle=0]{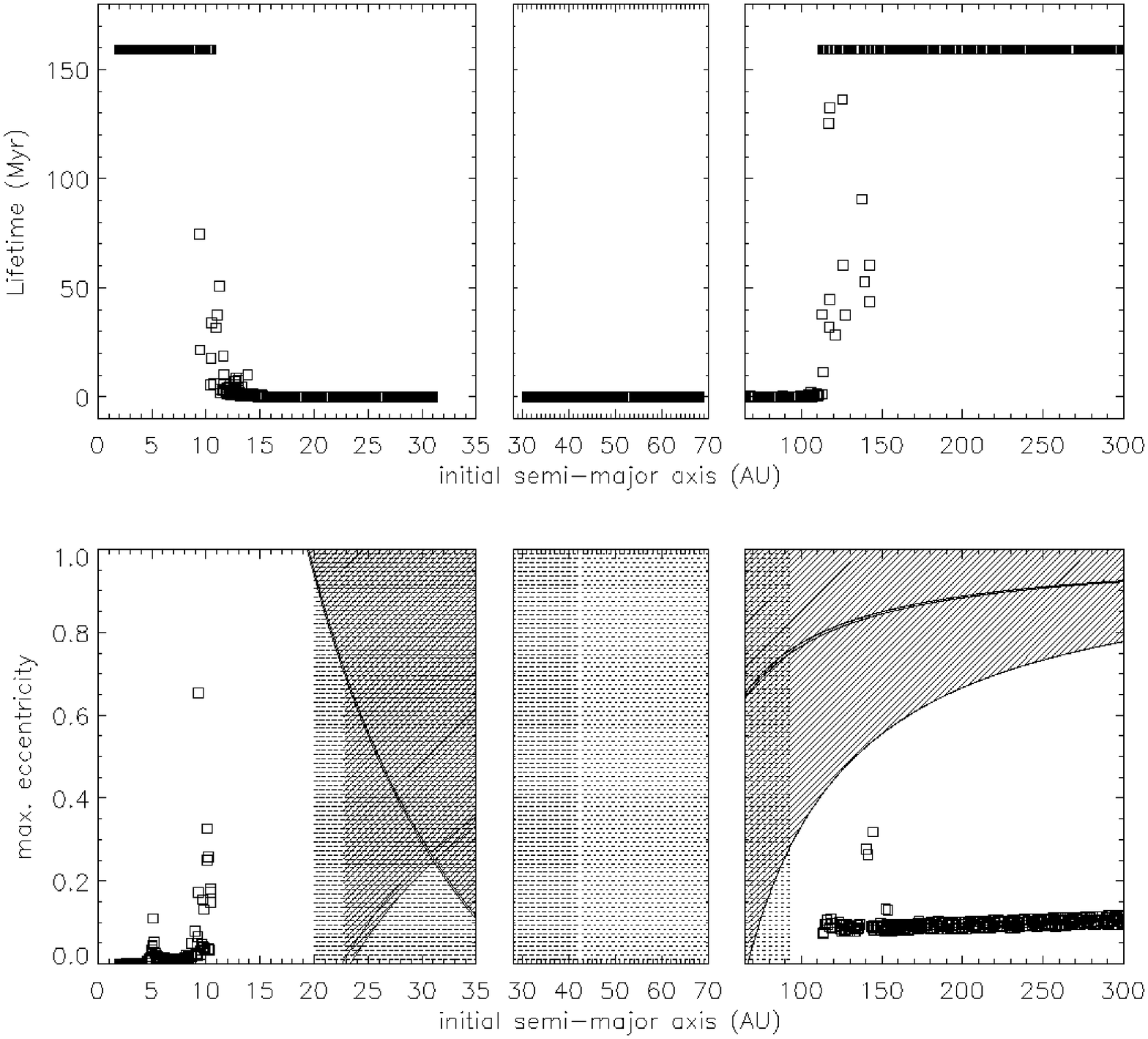}
\end{center}       
\caption{Same as Figure \ref{HR8799_dyn_D1} but for Fit D5 of HR 8799.
} 
\label{HR8799_dyn_D5}
\end{figure}

\begin{figure*}
\begin{center}
\includegraphics[scale=0.5,angle=0]{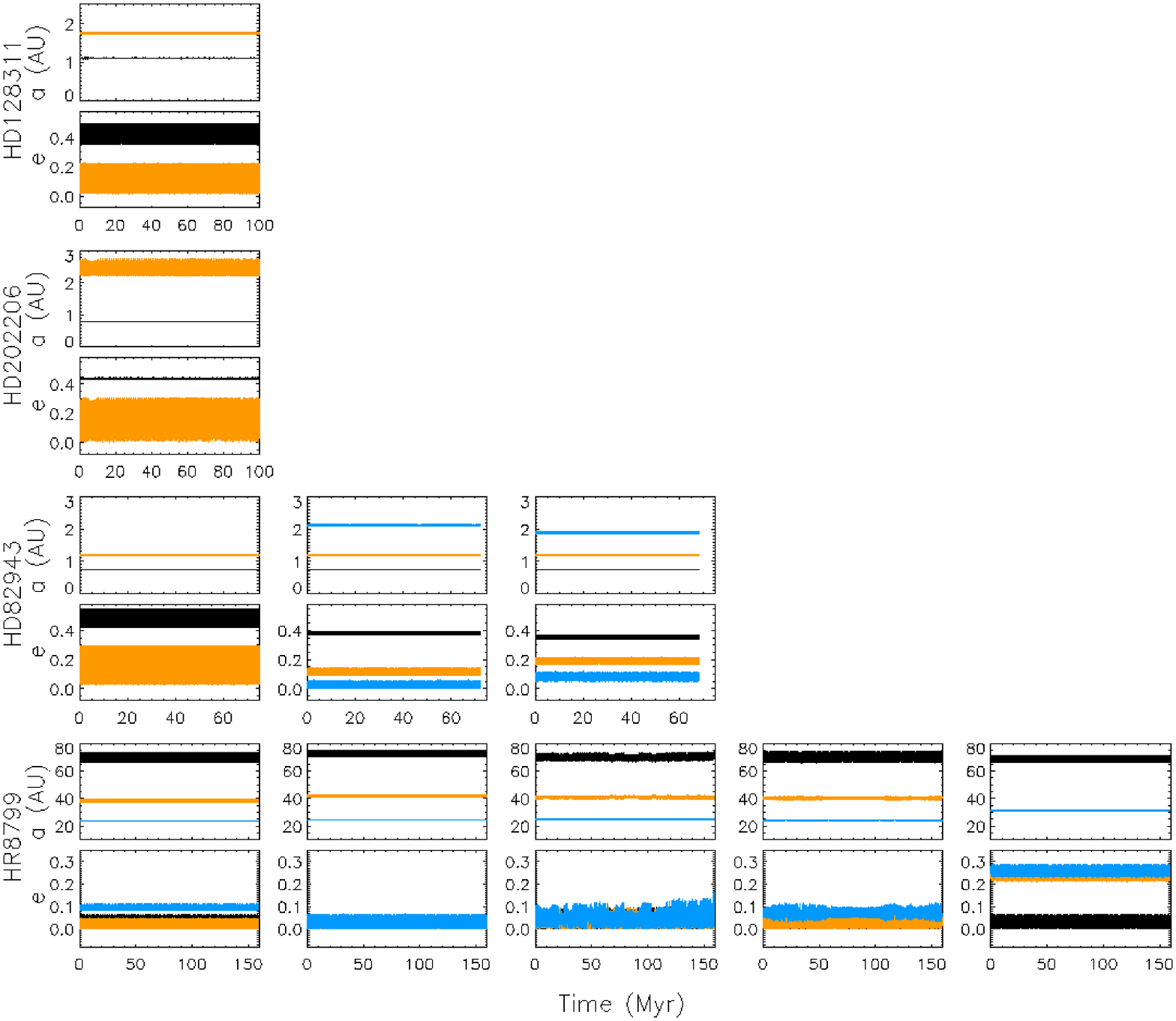}
\end{center}
\caption{Long-term evolution of the planets' semi-major axes and eccentricities for the planetary configurations in Table \ref{planetparam}. From top to bottom, left to right the panels correspond to fits A2, B1, C1--C3, D1--D5, for HD 128311, HD 202206, HD 82943 and HR 8799, respectively. The models were run for 100 Myr (HD 128311), 100 Myr  (HD 202206), $\gtrsim$70 Myr (HD 82943) and 160 Myr (HR 8799). The colors correspond to planet b (black), c (orange) and d (blue). 
} 
\label{ploe}
\end{figure*}

We have studied the stability of the potential dust-producing planetesimals using numerical simulations of test particles in the five planetary configurations discussed in Section \ref{obs_HR8799} and listed in Table \ref{planetparam}.  The simulations consisted of 1500 test particles: 500 particles uniformly spaced between 2 AU and the semi-major axis of the inner-most planet, 500 particles located between the innermost and outermost planet, and 500 particles uniformly spaced between the semi-major axis of the outermost planet and 300 AU. This latter value is taken from the results in the SED analysis in Su et al. (2009) summarized in Section \ref{sed_HR8799}. 
We assumed that the planets and the dust-producing planetesimals formed out of a thin disk and are co-planar.  Particles were removed if they approached the star closer than 2 AU, or approached a planet closer than its Hill radius. The orbits were integrated 
for 160 Myr (the upper limit to the stellar age estimated by Marois et al. 2008). The semi-major axes and eccentricities of the planet orbits do not evolve significantly during that time span (see  Figure \ref{ploe}).  The results regarding the test particles are shown in Figures \ref{HR8799_dyn_D1}--\ref{HR8799_dyn_D5}. For all the planetary configurations considered, the region between the orbits of the planets is dynamically unstable. The test particles' orbits are stable in two regions: (a) $\lesssim$ 12 AU for fits D1 and D2, and $\lesssim$ 10 AU  for fits D3, D4, D5; and (b) $\gtrsim$ 110 AU for fits D1, D2, D5, and $\gtrsim$ 150 AU for fits D3, D4.  Regarding the maximum eccentricity of the test particles in the dynamically stable regions, Figures \ref{HR8799_dyn_D1}--\ref{HR8799_dyn_D5} show that $e_{max}$ $\lesssim$ 0.3, even though for some fits there are peaks at $\sim$ 5 AU and $\sim$170 AU (due to secular resonances). 
Given the low maximum eccentricities, these dynamically stable regions could be possible locations of long-lived, dust-producing planetesimals (except at $\sim$ 5 AU in fit D3 where the $e_{max}$ reaches 0.4 and where the planetesimals lifetimes might be short under collisional evolution).

\vspace{1cm}
\section{Discussion: potential location of the dust-producing planetesimal belts}
\label{loc}   

We now discuss the potential location of the dust-producing planetesimal belts in each for the four multiple-planet systems studied in this paper. 

\subsection{HD 128311}
\label{loc_HD128311}
We favor  models that assume a single grain radius of 10 $\mu$m because they represent large blackbody grains in general, and there is no evidence of a significant population of small grains.  With these models, the observed SED can be fitted by a dust disk with R$_{out}$ = 10 AU and R$_{in}$ $\ge$ 5 AU, R$_{out}$ = 50 AU and R$_{in}$ $\ge$ 2 AU, or R$_{out}$ = 100 AU and R$_{in}$ $\ge$ 2 AU. 
The dynamical simulations help constrain further the possible location of the dust-producing planetesimals because the effects of the planets extend into the regions allowed by the SED models: they predict two stable niches where planetesimals could be long-lived, beyond  $\sim$ 4 AU and inside $\sim$ 0.3 AU. Regarding the latter, to account for the lack of excess emission at $\lambda$ $<$ 33 $\mu$m, the SED modeling excludes the presence of a significant population of dust-producing planetesimals in this region. To set tighter constraints to the location of the planetesimals in HD 128311 there is the need to obtain spatially resolved images and/or accurate photometric points in the 33 $\mu$m--70 $\mu$m range and in the sub-mm. As seen in Figures \ref{obs_3stars} and \ref{sed_hd128311}, observations with the recently launched {\it Herschel/PACS} would be very valuable for this purpose. 

\subsection{HD 202206}
\label{loc_HD202206}
The dynamical model indicates that planetesimals orbits are stable beyond $\sim$ 6 AU, and that their maximum eccentricity remains $\lesssim$ 0.1 for all semi-major axes, i.e. that planetesimals could be long-lived beyond 6 AU. This result agrees with the conclusions from the SED modeling. These models, which are well-constrained because of the presence of a small excess beyond 25 $\mu$m, result in a relatively narrow range of planetesimal belts that can fit the observations, ranging from a 50 AU  disk with an inner cavity 10--20 AU in size, to a 100 AU  disk with an inner cavity 6--10 AU in size.  We conclude that the gravitational perturbations of the detected planets might be responsible for the inner edge of the dust disk.     

\subsection{HD 82943}
\label{loc_HD82943}
The observed SED, lacking emission at $\lambda$ $<$ 33 $\mu$m, can be fitted by a dust disk composed of single grains 10 $\mu$m in size, with an inner cavity with R$_{in}$ $\ge$16 AU for a disk with R$_{out}$ = 50 AU, or  R$_{in}$ = 12 --26 AU for a disk with R$_{out}$ = 100 AU (a compact 10 AU dust disk is excluded). Because the dynamical modeling of test particles in the three planetary configurations considered suggests that planetesimals could be stable and long-lived beyond $\sim$ 3 AU,  we conclude that the gravitational perturbations from the planets (located within 2.1 AU of the star) do not extend far enough to have a significant effect to destabilize any debris system that is seen in the infrared excess emission. 

\subsection{HR 8799}
\label{loc_HR8799}
\begin{figure*}
\begin{center}
\includegraphics[scale=0.9,angle=0]{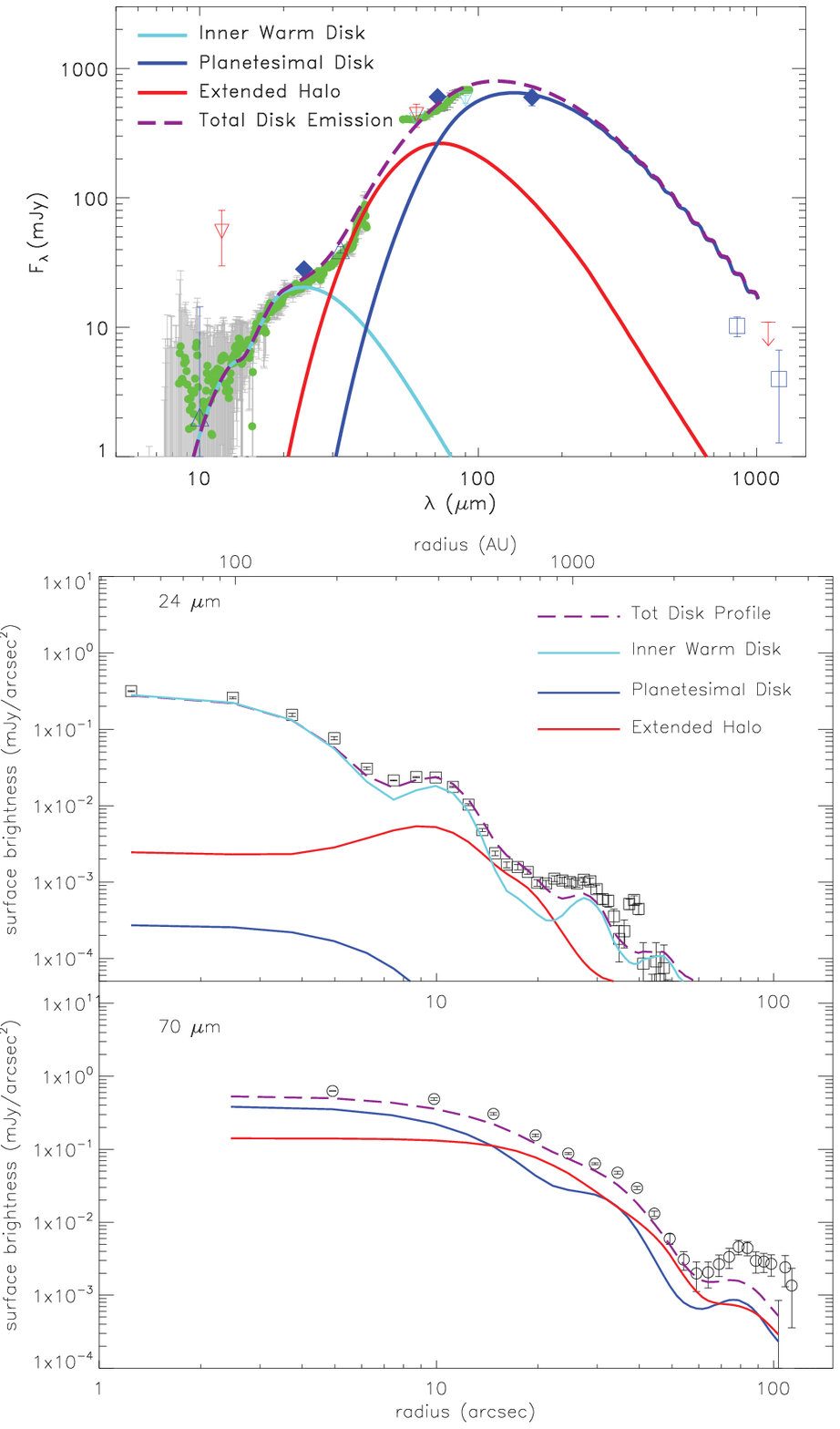}
\end{center}
\caption{Same as Figure 9 in Su et al. (2009) but setting the inner edge of the cold component at R$_{in}$ = 110 AU instead of 90 AU. Compared to the model with R$_{in}$ = 90 AU, this fit gives a slightly worse $\chi$$^2$-value (by 20\%) but is still within the errors.  (Top) Observed and modeled SED: the observations are represented as symbols with error bars and correspond to MIPS (dark blue diamonds; color-corrected), MIPS-SED (green), IRS (green), ISO (light blue triangles; color corrected), IRAS (red triangles; color corrected) and sub-millimeter (dark blue squares); regarding the latter, even though it appears that the model overestimates the flux in the sub-millimeter, both fluxes agree when taking into account the source extension (see Su et al. 2009 for details).  (Bottom) Observed and modeled surface brightness profiles at 24 $\mu$m and 70 $\mu$m; the observations are represented as open symbols with error bars. 
The parameters of the three disk components are as follows. Inner warm disk: R$_{in}$ = 6 AU, R$_{out}$ = 16 AU, a flat surface density ($\Sigma$ $\sim$ $r^{0}$), M$_{dust}$ = 1.12$\cdot$10$^{-6}$ M$_{\oplus}$, and a grain size distribution with $b_{min}$ = 1.5 $\mu$m, $b_{max}$ = 4.5~$\mu$m and $q$=-3.5.  Outer cold disk:  R$_{in}$ = 110 AU, R$_{out}$ = 300 AU,  a flat surface density ($\Sigma$ $\sim$ $r^{0}$), M$_{dust}$ = 1.31$\cdot$10$^{-1}$ M$_{\oplus}$, and a grain size distribution with $b_{min}$ = 10 $\mu$m, $b_{max}$ = 1000~$\mu$m and $q$=-3.5.  Extended halo:  R$_{in}$ = 300 AU, R$_{out}$ = 1000 AU,  a surface density given by $\Sigma$ $\sim$ $r^{-1}$, M$_{dust}$ = 1.94$\cdot$10$^{-2}$ M$_{\oplus}$, and a grain size distribution with $b_{min}$ = 1 $\mu$m, $b_{max}$ = 10~$\mu$m and $q$=-3.5. See Su et al. (2009) for details on the modeling. 
} 
\label{HR8799_sed_new}
\end{figure*}

For the five planetary configurations considered, the dynamical modeling of test particles suggests that planetesimals could be stable and long-lived at semi-major axes $\lesssim$ 10--12AU and $\gtrsim$ 110-150 AU. This helps to further constrain the dust disk solutions based on the SED fitting by Su et al. (2009), who proposed: (1) the presence of an unresolved warm disk 15 AU in size, if the grain sizes range from 1 to 4.5 $\mu$m, or a more compact 10 AU disk, if  all the grains are bound ($>$ 2 $\mu$m); and (2) the presence of an unresolved cold disk with an inner edge at $\sim$ 90 AU. The dynamical models favor an inner dust disk component of bound grains and R$_{out}$ $\sim$ 10 AU and a colder component with R$_{in}$ $\sim$ 110--150 AU (rather than $\sim$ 90 AU). Regarding the cold component, we find there is no tension between R$_{in}$ derived from the dynamical models ($\sim$ 110--150 AU) and that derived from the analysis of the SED and the surface brightness radial profiles. Even though Su et al. (2009) favors a value of R$_{in}$ $\sim$ 90 AU,  Figure \ref{HR8799_sed_new} shows that the observations can also be fitted with R$_{in}$ = 110 AU (increasing the total dust mass) and R$_{in}$ = 150 AU (changing the minimum grain size from 10 $\mu$m to 8 $\mu$m); this is not surprising because all these inner radii are smaller than the $MIPS$ 70 $\mu$m pixel size. 

The presence of the outer planetesimal disk may help constrain planet formation scenarios that have been proposed for the HR8799 system. One scenario is planet-planet scattering: in addition to the difficulty to result in a stable system with low eccentric planets (Dodson-Robinson et al. 2009), this model may not be able to account for the presence of the outer planetesimal disk. A second scenario is  long-ranged outward migration in resonance (Crida et al. 2009): future high-resolution  imaging of the cold component of the HR 8799 debris disk may be able to provide evidence of dust-producing planetesimals trapped in MMRs with the outermost planet, a signpost of outward planet migration. 

Table \ref{multidisks} and Figure \ref{arch} summarize the possible planet-planetesimals configurations of HD 128311, HD 202206, HD 82943 and HR 8799, compared to that of the other three multi-planet systems known to harbor dust -- HD 38529, HD 69830 and the Sun. In some cases, the SED allows for the presence of both narrow and wide belts. We favor the latter because the narrowest belts might only be justified in the presence of additional planetary perturbers for which we have no evidence so far. However,  one should keep in mind the degeneracy: to set tighter constraints to the location of the planetesimals there is the need to obtain spatially resolved images and/or accurate photometric points in the 30 $\mu$m--70 $\mu$m range and in the sub-mm. Observations with {\it Herschel/PACS} and ALMA will be very valuable for this purpose. 

\section{Conclusions}

In this paper we have studied the possible planet-planetesimal
configurations of four multi-planet systems, of which three are
radial-velocity systems -- HD 128311, HD 202206, HD 82943 -- and one is a directly imaged system -- HR 8799. We have quantified where the zone of influence lies of planets on the dust-producing planetesimals: for HR 8799 it extends to nearly 20 AU from the orbit of the outermost planet, while for the three radial-velocity systems it extends to about 4 AU.  A previous paper that studied HD38529, another radial-velocity multi-planet system, found that the influence of the planets in this case extends out to $\sim$ 10 AU (determining the inner edge of the disk), and becomes dominant again at a $\sim$ 55 AU (due to a secular resonance that probably determines the outer edge of the dust disk (Moro-Mart\'in et al. 2007b). We conclude that radial-velocity multi-planet systems generally have zones of influence within a few to ten AU; more precise determinations will require individual modeling of a system. The influence can be extended much further through secular resonances and similar behavior.  

For the three radial-velocity multi-planet systems studied in this paper, we have constructed fits to the spectral energy distributions of the debris disks. If we use astronomical silicates with a size distribution down to the blow-out size, the emitting regions of the disks must be so far from the star that they are well outside the zones of influence of the radial velocity planets. This behavior depends critically on the optical properties of the grains (size distribution and optical constants of the grain material). If we were to adopt optical constants typical of an ice-silicate mixture instead of astronomical silicates, the dust may be located closer to the star. In all three cases, we find that single-size 10 $\mu$m astronomical silicates reproduce well the observed SED, and could lie at the edge of the zone of influence of the planets and therefore the disk would be sculpted by them.


\begin{center} {\it Acknowledgments} \end{center}
We thank Hal Levison for providing  skeel-SyMBA  for the dynamical simulations, Sebastian Wolf for providing DDS for the SED models, Alexander Krivov for his careful reading of the manuscript and useful comments and Daniel Fabrycky and Ruth Murray-Clay for useful comments. This work is based on observations made with the {\it Spitzer} Space Telescope,
which is operated by the Jet Propulsion Laboratory, managed for NASA by the  California Institute of Technology. 
A.M.M. acknowledges funding from the Spanish MICINN (Ram\'on y Cajal Program and grants AYA2009-07304 and CONSOLIDER INGENIO 2010CSD2009-00038), the Michelson Fellowship and the {\it Spitzer} archival grant 40412. She thanks the Isaac Newton Institute for Mathematical Sciences at Cambridge University for support. RM acknowledges support from grants by NSF (AST-0806828) and NASA (NNX08AQ65G).

\clearpage
\LongTables
\begin{landscape}
\begin{deluxetable}{lllllllllllllllllllc}
\tiny
\tablewidth{120pc}
\tablecaption{Stars with evidence of harboring a multiplanet system and planetesimals\label{multidisks}} 
\tablehead{
\colhead{Source}	& {SpTyp} 	& {Excess $\lambda$ ($\mu$m)} 		& \multicolumn{4}{c}{Planetesimals' Location\tablenotemark{a} (AU)} 								& \multicolumn{11}{c}{Planetary Configuration\tablenotemark{b}}\\
\colhead{} 		& {Age (Gyr)}	& {L$_{dust}$/L$_{*}$(10$^{-4}$)} 		& \multicolumn{2}{c}{10$\mu$m grains} 			& \multicolumn{2}{c}{size distribution} 						&						&\multicolumn{4}{c}{Fit 1}			&\multicolumn{3}{c}{Fit 2}		&\multicolumn{3}{c}{Fit 3}		\\	
\colhead{} 		& {}			& {} 								&			&							&			&										&						&\multicolumn{4}{c}{Planet} 		&\multicolumn{3}{c}{Planet}		&\multicolumn{3}{c}{Planet}	\\
\colhead{} 		& {}			& {} 								& {R$_{in}$} 	& {R$_{out}$}	 				& {R$_{in}$} 	& {R$_{out}$}								&						&{b}&{c}&{d}&{e}				&{b}&{c}&{d}					&{b}&{c}&{d}				}
\startdata

HD 128311   		& K0 V		& 70								& $\ge$5			& 10\tablenotemark{e} 			 			& $\ge$52		& 100\tablenotemark{e} 				&$M$					& 2.19 & 3.22 &	 \nodata & \nodata	& 1.61 & 3.18 \\
				& 0.5			& 0.13--0.27 						& $\ge$2			& 50\tablenotemark{e} 						& 				&										&$a$					& 1.10 & 1.76 &	 \nodata & \nodata	& 1.11 & 1.73 \\
				&				&									& $\ge$2			& 100\tablenotemark{e} 						&				&										&$e$					& 0.25 & 0.17 &	 \nodata & \nodata	& 0.36 & 0.21 \\
\hline\\				
HD 202206 	& G6 V& 25--70				& 10--20			& 50											& 				& 										&$M$					& 16.6 & 2.19 & \nodata & \nodata \\				
 									& 5.6	& 						& 6	--10			& 100											&				&										&$a$					& 0.80 & 2.51 &	 \nodata & \nodata \\				
 									& 		& 						& 					&	 											&				&										&$e$					& 0.44 & 0.07 &	 \nodata & \nodata \\			
\hline\\				
HD 82943		         				& G0 V& 70					& 16				& 50\tablenotemark{e}						&				& 										&$M$					& 1.46 & 1.73  & \nodata & \nodata 	 		& 1.68 & 1.87 & 0.49 			& 1.70 & 1.74 & 0.35 \\		
									& 5	& 0.88--1				& 12--26			& 100	    										&				&										&$a$ 					& 0.75 & 1.19  & \nodata & \nodata   	 		& 0.75 & 1.20 & 2.12    			& 0.74 & 1.20 & 1.19 \\	
									&		&						&					&												&				&										&$e$					& 0.45 & 0.27  & \nodata & \nodata   		     	& 0.39 & 0.11 & 0.018 		& 0.36 & 0.19 & 0.078\\
\hline\\			
HR 8799\tablenotemark{c}& A5 V	& 8--850 							&  \multicolumn{4}{c}{warm: 6 to 10}     																&$M$					& 7 	   & 10      & 10          & \nodata	& 8.02 & 11.9 & 8.89 		& 9.71 & 7.96 & 7.40 \\		
				& 0.03--0.16	& 2.3 			       				&  \multicolumn{4}{c}{cold: 110--150 to 300}															&$a$					& 67.1 & 38.0  & 23.4       & \nodata	& 68.4 & 39.6 & 24.2                   & 67.7 & 31.0 &  30.8 \\
				&			&								&			&							&			&										&$e$					& 0 	   & 0        & 0.09       & \nodata	&0.01 & 0.01 & 0.07		         & 0.01 & 0.25 & 0.27  \\
\hline\\				
HD 38529 		& G8 III/IV 		& 70								& 15			& 50							&			&										&$M$					& 0.85 & 13.2  	& \nodata & \nodata  \\
				& 3.5	 		& 0.36							& 8			& 100						&			&										&$a$					& 0.13 &  3.74 	& \nodata & \nodata  \\
				&			&								& R$_{sub}$     & 500						&			&										&$e$					& 0.25 & 0.35    & \nodata & \nodata  \\				
\hline\\				
HD 69830   		& K0 V		& 8--35							& \multicolumn{4}{c}{0.93--1.16}   																	&$M$					& 0.03 & 0.04   & 0.06       &  \nodata \\
				& 4--10		& 2 								&			&							&		         &										&$a$					& 0.08 & 0.19   & 0.63 	& \nodata \\
				& 			&  								&			&							&		         &										&$e$					& 0.10 & 0.13  & 0.07  	& \nodata \\
\hline\\				
Sun\tablenotemark{d}& G2 V		& (0.1--1)$\cdot$10$^{-3}$ 			& \multicolumn{4}{c}{2--4 (Asteroid Belt)}   															&$M$					& 1	   & 0.30	& 0.05 & 0.05 \\
				& 4.57		& (1--10)$\cdot$10$^{-3}$ 			& \multicolumn{4}{c}{35--50 (Kuiper Belt)}   															&$a$					&  5.20 & 9.58	& 19.2 & 	30.1 \\
				&			&  								& 			& 							&			&										&$e$					& 0.05 &0.05	& 0.04  &  0.01\\
\enddata
\tablenotetext{a}{Inferred planetesimals' location. HD 128311: this work; HD 202206: this work; HD 82943: this work; HR 8799 : Su et al. (2009) and this work;  HD 38529: Moro-Martin et al. 2007b; HD 69830: Lisse et al. 2007.}
\tablenotetext{b}{Planetary parameters: $M$, $a$ and $e$ are planet mass in M$_{Jup}$, semi-major axis in AU and eccentricity, respectively. The rest of the orbital elements are listed in Table \ref{planetparam}.}
\tablenotetext{c}{For HR 8799, this table only includes the planetary configurations where the planets have their nominal masses; the SED reveals the presence of a warm component and a cold component of the dust disk (from Su et al. 2009).}
\tablenotetext{d}{Fractional luminosities of asteroidal dust and Kuiper belt dust from Dermott et al. (2002) and Stern (1996).}
\tablenotetext{e}{Narrow belts of width 10\% of the disk radius give valid fits to the SED (but see discussion in Section \ref{sed_HD128311}).}
\end{deluxetable}
\clearpage
\end{landscape}

\newpage
\begin{thebibliography}{}
\bibitem{2009ApJ...690.1522B} 
Backman, D., et al.\ 2009,  ApJ, 690, 1522 
\bibitem{2007ApJ...669.1167B} 
Barnes, S.~A.\ 2007,  \apj, 669, 1167 
\bibitem[Beaug\'e et al. (2008)]{beau08}
Beaug\'e, C., Giuppone, C. A., Ferraz-Mello, S. \& Michtchenko, T. A. 2008, MNRAS, 385, 2151
\bibitem[Beichman et al. (2005)]{beic05}
Beichman, C. A. et al. 2005, ApJ, 622, 1160
\bibitem{2006ApJ...639.1166B} 
Beichman, C.~A., et al.\ 2006,  \apj, 639, 1166
\bibitem[Biller et al. (2007)]{bill07} 
Biller, B. A.  et al. 2007, ApJS, 173, 143
\bibitem[Bryden et al.(2009)]{2009ApJ...705.1226B} 
Bryden, G., et al.\ 2009, \apj, 705, 1226 
\bibitem[Butler (2006)]{butl06}
Butler, R.P. et al. 2006, ApJ, 646, 505
\bibitem{1979Icar...40....1B} 
Burns, J.~A., Lamy, P.~L., \& Soter, S.\ 1979,  Icarus, 40, 1 
\bibitem[Carpenter et al. (2009)]{carp09} 
Carpenter, J. M. et al. 2009, ApJS, 181, 197
\bibitem[Chauvin et al.(2006)]{2006A&A...456.1165C} 
Chauvin, G. et al. 2006, \aap, 456, 1165 
\bibitem[Chen et al.(2009)]{2009ApJ...701.1367C} 
Chen, C.~H. et al. 2009, ApJ, 701, 1367 
\bibitem[Correia et al.(2005)]{2005A&A...440..751C} 
Correia, A.~C.~M. et al. 2005, \aap, 440, 751 
\bibitem[Couetdic et al.(2009)]{2009arXiv0911.1963C} 
Couetdic, J. et al. 2009, arXiv:0911.1963 
\bibitem{2009ApJ...705L.148C} Crida, A., Masset, F., 
\& Morbidelli, A.\ 2009,  ApJL, 705, L148 
\bibitem{2002ESASP.500..319D} 
Dermott, S.~F., Kehoe, T.~J.~J., Durda, D.~D., Grogan, K., \& Nesvorn{\'y}, D.\ 2002,  Asteroids, Comets, and Meteors: ACM 2002, 500, 319 
\bibitem{2009ApJ...707...79D} Dodson-Robinson, S.~E., Veras, D., Ford, 
E.~B., \& Beichman, C.~A.\ 2009,  ApJ, 707, 79 
\bibitem[Duncan et al.(1998)]{1998AJ....116.2067D} 
Duncan, M.~J., Levison, H.~F., \& Lee, M.~H.\ 1998, AJ, 116, 2067 
\bibitem[Eggenberger et al. (2007)]{egge07} 
Eggenberger, A. et al. 2007, A\&A, 474, 273
\bibitem[Fabrycky \& Murray-Clay(2010)]{2010ApJ...710.1408F} Fabrycky, D.~C., 
\& Murray-Clay, R.~A.\ 2010, ApJ, 710, 1408 
\bibitem[Ferraz-Mello et al.(2005)]{2005ApJ...621..473F} 
Ferraz-Mello, S., Michtchenko, T.~A., \& Beaug{\'e}, C.\ 2005, ApJ, 621, 473 
\bibitem{2009ApJ...697.1578G} 
G{\'a}sp{\'a}r, A., Rieke, G.~H., Su, K.~Y.~L., Balog, Z., Trilling, D., Muzzerole, J., Apai, D., 
\& Kelly, B.~C.\ 2009,  \apj, 697, 1578 
\bibitem{2003AJ....126.2048G} 
Gray, R.~O., Corbally, C.~J., Garrison, R.~F., McFadden, M.~T., \& Robinson, P.~E.\ 2003,  \aj, 126, 2048 
\bibitem{2006AJ....132..161G} 
Gray, R.~O., Corbally, C.~J., Garrison, R.~F., McFadden, M.~T., Bubar, E.~J., McGahee, C.~E., O'Donoghue, A.~A., \& Knox, E.~R.\ 2006,  \aj, 132, 161 
\bibitem[Go{\'z}dziewski et al.(2006)]{2006ApJ...645..688G} 
Go{\'z}dziewski, K., Konacki, M., \& Maciejewski, A.~J.\ 2006, \apj, 645, 688 
\bibitem[Go{\'z}dziewski \& Konacki(2006)]{2006ApJ...647..573G} 
Go{\'z}dziewski, K., \& Konacki, M.\ 2006, ApJ, 647, 573 
\bibitem[Go{\'z}dziewski \& Migaszewski(2009)]{2009MNRAS.397L..16G} 
Go{\'z}dziewski, K., \& Migaszewski, C.\ 2009, MNRAS, 397, L16 
 \bibitem{2010AAS...21537707K} 
 Kalas, P., et al.\ 2010,  Bulletin of the American Astronomical Society, 41, 587 
\bibitem{2003AJ....125.1980K} 
King, J.~R., Villarreal, A.~R., Soderblom, D.~R., Gulliver, A.~F., \& Adelman, S.~J.\ 2003,  \aj, 125, 1980 
\bibitem{2006A&A...455..509K} Krivov, A.~V., L{\"o}hne, T., \& Srem{\v c}evi{\'c}, M.\ 2006,  A\&A, 455, 509 
\bibitem{2010RAA....10..383K} Krivov, A.~V.\ 2010,  Research in Astronomy 
and Astrophysics, 10, 383 
\bibitem[Hillenbrand et al. (2008)]{hill08} 
Hillenbrand, L. A. et al. 2008, ApJ, 677, 630
\bibitem{2009ApJ...705...89L} 
Lawler, S.~M., et al.\ 2009,  \apj, 705, 89 
\bibitem[Lafreni{\`e}re et al.(2009)]{2009ApJ...694L.148L} 
Lafreni{\`e}re, D., Marois, C., Doyon, R., \& Barman, T.\ 2009, ApJL, 694, L148 
\bibitem[Lee et al.(2006)]{2006ApJ...641.1178L} 
Lee, M.~H., Butler, R.~P., Fischer, D.~A., Marcy, G.~W., \& Vogt, S.~S.\ 2006, ApJ, 641, 1178 
\bibitem[Lisse et al. (2007)]{liss07} 
Lisse, C. M., Beichman, C. A., Bryden, G. \& Wyatt, M. C. 2007, ApJ, 658, 584
\bibitem[Lovis et al. (2006)]{lovi06} 
Lovis, C. et al. 2006, Nature, 441, 305
\bibitem[Nelson(2003)]{2003MNRAS.345..233N} 
Nelson, R.~P.\ 2003, \mnras, 345, 233 
\bibitem{2008ApJ...687.1264M} 
Mamajek, E.~E., \& Hillenbrand, L.~A.\ 2008,  \apj, 687, 1264 
\bibitem[Marois et al. 2008]{maro08} 
Marois, C. et al. 2008, Science, 322, 1348
\bibitem[Mayor et al.(2004)]{2004A&A...415..391M} 
Mayor, M. et al.\ 2004, A\&A, 415, 391 
\bibitem[Metchev et al.(2009)]{2009ApJ...705L.204M} 
Metchev, S., Marois,  C., \& Zuckerman, B.\ 2009, ApJL, 705, L204 
\bibitem{2008ApJ...673L.181M} 
Meyer, M.~R., et al.\ 2008,  \apjl, 673, L181
\bibitem[Mo{\'o}r et al.(2006)]{2006ApJ...644..525M} 
Mo{\'o}r, A.,  A. et al. \ 2006, ApJ, 644, 525 
\bibitem[Moro-Mart\'{\i}n, Wolf \& Malhotra (2005)]{ama05a}
Moro-Mart\'{\i}n, A., Wolf, S. \& Malhotra, R. 2005, ApJ, 621, 1079
\bibitem[Moro-Mart\'{\i}n et al. (2007)]{ama07a} 
Moro-Mart\'{\i}n, A. et al. 2007, ApJ, 658, 1312
\bibitem[Moro-Mart\'{\i}n et al. (2007)]{ama07b} 
Moro-Mart\'{\i}n, A. et al. 2007, ApJ, 668, 1165
\bibitem[Moro-Mart\'{\i}n et al. (2008)]{2008ssbn.book..465M} 
Moro-Mart\'{\i}n, A., Wyatt, M.~C., Malhotra, R., \& Trilling, D.~E.\ 2008, in The Solar System Beyond Neptune (M. A. Barucci, H. Boehnhardt, D. P. Cruikshank, and A. Morbidelli eds.), University of Arizona Press, Tucson, p.465-480. 
\bibitem{2010MNRAS.tmpL..61M} 
Moya, A., Amado, P.~J., Barrado, D., Garc{\'{\i}}a Hern{\'a}ndez, A., Aberasturi, M., Montesinos, B., \& Aceituno, F.\ 2010,  MNRAS, L61
\bibitem{2010ApJ...708.1728M} M{\"u}ller, S., L{\"o}hne, T., 
\& Krivov, A.~V.\ 2010,  ApJ, 708, 1728 
\bibitem[Reidemeister et al.(2009)]{2009A&A...503..247R} 
Reidemeister, M. et al. 2009, A\&A, 503, 247 
\bibitem[Saffe, G\'omez \& Chavero (2005)]{saff05}
Saffe, C., G\'omez, M. \& Chavero, C. 2005, A\&A, 443, 609
\bibitem{saff08} 
Saffe, C., G\'omez, M., Pintado, O. \& Gonz\'alez, E. 2008, A\&A, 490, 297
\bibitem{2010ApJ...712.1421S} 
Sierchio, J.~M., Rieke, G.~H., Su, K.~Y.~L., Plavchan, P., Stauffer, J.~R., \& Gorlova, N.~I.\ 2010,  \apj, 712, 1421 
\bibitem[Sousa et al.(2008)]{2008A&A...487..373S} 
Sousa, S.~G., et al.\ 2008, A\&A, 487, 373 
\bibitem[Stern (1996)]{ster96}
Stern, S. A. 1996, AJ, 112, 1203
\bibitem[Su et al.(2009)]{2009ApJ...705..314S} 
Su, K.~Y.~L., et al.\ 2009, ApJ, 705, 314 
\bibitem[Sylvester et al.(1996)]{1996MNRAS.279..915S} 
Sylvester, R.~J., Skinner, C.~J., Barlow, M.~J., \& Mannings, V.\ 1996, MNRAS, 279, 915 
\bibitem{2007A&A...472..169T} Th{\'e}bault, P., \& Augereau, J.-C.\ 2007,  A\&A, 472, 169 
\bibitem[Trilling et al. (2008)]{tril08} 
Trilling, D. E. et al. 2008, ApJ, 674, 1086
\bibitem[Udry et al.(2001)]{2001yCat..33560590U} 
Udry, S., et al.\ 2001, VizieR Online Data Catalog, 335, 60590 
\bibitem[Valenti \& Fischer (2005)]{vale05}
Valenti, J. A. \& Fischer, D. A. 2005, ApJS, 159, 141
\bibitem[Vogt et al. (2005)]{2005ApJ...632..638V} 
Vogt, S.~S. et al. 2005, ApJ, 632, 638 
\bibitem[Wisdom (1980)]{1980AJ.....85.1122W} 
Wisdom, J. 1980, AJ, 85, 1122 
\bibitem[Weingartner \& Draine (2001)]{wein01}
Weingartner, J.C. \& Draine, B. T. 2001, ApJ, 548, 296 
\bibitem[Williams \& Andrews(2006)]{2006ApJ...653.1480W} 
Williams, J.~P., \& Andrews, S.~M.\ 2006, ApJ, 653, 1480 
\bibitem{2004ApJS..152..261W} 
Wright, J.~T., Marcy, G.~W., Butler, R.~P., \& Vogt, S.~S.\ 2004,  \apjs, 152, 261 
\bibitem[Wolf \& Hillenbrand (2003)]{wolf03}
Wolf, S. \& Hillenbrand, L.A. 2003, ApJ, 596, 603
\end{thebibliography}
\end{document}